\documentclass[twocolumn,superscriptaddress]{aastex63}

\accepted{to PSJ Dec 19, 2023}

\usepackage{graphicx}
\usepackage{bm}
\usepackage{mhchem}
\usepackage{longtable}
\usepackage{supertabular}
\usepackage{textpos}
\usepackage{amsmath}
\usepackage{soul}
\usepackage{textcomp}

\usepackage{wasysym}

\newcommand{\beginappendixA}{%
        \setcounter{table}{0}
        \renewcommand{\thetable}{A\arabic{table}}%
        \setcounter{figure}{0}
        \renewcommand{\thefigure}{A\arabic{figure}}%
        \setcounter{equation}{0}
        \renewcommand{\theequation}{A\arabic{equation}}%
     }

\shorttitle{Organic hazes as a source of life's building blocks}
\shortauthors{Pearce et al.}

\begin{document}

\title{Organic hazes as a source of life's building blocks to warm little ponds on the Hadean Earth}

\author{Ben K. D. Pearce*}
\affiliation{Department of Earth and Planetary Science, Johns Hopkins University, Baltimore, MD, 21218, USA}
\thanks{bpearce6@jhu.edu; chaohe23@ustc.edu.cn}

\author{Sarah M. H{\"o}rst}
\affiliation{Department of Earth and Planetary Science, Johns Hopkins University, Baltimore, MD, 21218, USA}

\author{Joshua A. Sebree}
\affiliation{Department of Chemistry and Biochemistry, University of Northern Iowa, Cedar Falls, IA, USA}

\author{Chao He*}
\affiliation{School of Earth and Space Sciences, University of Science and Technology of China, Hefei, China}
\affiliation{Department of Earth and Planetary Science, Johns Hopkins University, Baltimore, MD, 21218, USA}

\begin{abstract}
{\bf 
Over 4 billion years ago, Earth is thought to have been a hazy world akin to Saturn's moon Titan. The organic hazes in the atmosphere at this time could contain a vast inventory of life's building blocks, and thus may have seeded warm little ponds for life. In this work, we produce organic hazes in the lab in atmospheres with high (5\%) and low (0.5\%) \ce{CH4} abundances and analyze the solid particles for nucleobases, amino acids, and a few other organics using GC/MS/MS to obtain their concentrations. We also analyze heated (200 $^{\circ}$C) samples from the high methane organic haze experiment to simulate these particles sitting on an uninhabitable surface. Finally, we use our experimental results and estimates of atmospheric haze production as inputs for a comprehensive numerical pond model to calculate the concentrations of nucleobases from organic hazes in these environments. We find that organic hazes typically provide up to 0.2--6.5 $\mu$M concentrations of nucleobases to warm little ponds for potentially habitable Hadean conditions. However, without seepage, uracil and thymine can reach $\sim$100 $\mu$M concentrations, which is the present lower experimental limit to react these species to form nucleotides. Heating samples leads to partial or complete decay of biomolecules, suggesting that biomolecule stockpiling on the hot surface is unlikely. The ideal conditions for the delivery of life's  building blocks from organic hazes would be when the Hadean atmosphere is rich in methane, but not so rich as to create an uninhabitable surface.
}
\end{abstract} 


\section*{Introduction}

Similar to present day Titan, the early Hadean Earth ($\sim$4.5--4.3 billion years ago) is thought to have been a hazy world due to its hydrogen- and methane-rich atmosphere \citep{Pavlov_et_al2001,Trainer_et_al2004,Trainer_et_al2006,DeWitt_et_al2009,Reference124,Horst_et_al2018}.
Organic hazes produced in these conditions contain biomolecules such as the nucleobases that make up ribonucleic acid (RNA), as well as the amino acids that make up proteins \citep{Kawai_et_al2019,Sebree_et_al2018,2012AsBio..12..809H,Neish_et_al2010}. We ask, could these organic hazes have seeded early Earth for life?

The organic haze particles that would have been useful for prebiotic chemistry would be those that landed in small bodies of water such as ponds. Ponds are favorable sites for the origin of life as their periodic evaporation-rehydration cycles provide the conditions necessary to concentrate biomolecues and polymerize RNA \citep{Damer_Deamer2019,Reference83}. Other sources of biomolecules to ponds have been studied, including meteorites, interplanetary dust, and hydrogen cyanide (HCN) chemistry within the pond \citep{Reference425,Pearce_et_al2022a}. Meteorites offer the highest concentrations of RNA building blocks to ponds, at $\sim$10 $\mu$M; however, this finite source depletes in weeks \citep{Reference425}. HCN by contrast is constantly raining out of the troposphere, but only provides sub-$\mu$M concentrations of RNA building blocks in scenarios where atmospheric methane is sourced from hydrothermal systems \citep{Pearce_et_al2022a}. Such concentrations are at least a couple orders of magnitude lower than any experiment performed in the lab that reacts these species to form nucleosides (ribose + nucleobase) or nucleotides (ribose + nucleobase + phosphate) \citep{1963Natur.199..222P}.

Hydrogen is thought to be the dominant atmospheric species ($\sim$60--90\%) at the earliest stages of the Hadean as it is produced rapidly when iron delivered from impactors reacts with surface water (\ce{Fe + H2O -> FeO + H2}) \citep{Pearce_et_al2022a,Zahnle_et_al2020}. Methane can be produced in abundances of 0.1--10\% from the heat generated by impactors 400--500 km in size - similar to the asteroid Vesta \citep{Zahnle_et_al2020}. Molecular nitrogen and carbon dioxide would have also been present from volcanic outgassing \citep{Reference119}.

Although hydrogen and methane-rich atmospheres are favorable for organic haze production, too much of these species in the early atmosphere would inhibit ponds from forming on the surface. This is because both methane, and \ce{H2}-\ce{H2} (that forms at $>$ 1 bar atmospheric pressures) are potent greenhouse gases. It is unclear at which atmospheric pressures of \ce{CH4} the early Earth atmosphere would become uninhabitable; however, some unrefined radiative transfer climate models of reducing Hadean atmospheres produce surface temperatures of $\sim$110--350 $^{\circ}$C for methane abundances around 5\% \citep{Cerrillo_et_al2023}. On the other hand, these models suggest that an early Earth atmosphere with $\sim$0.5\% methane may support cool enough temperatures for ponds to exist on the surface \citep{Cerrillo_et_al2023}; thus, we explore these two concentrations in our experiments and define them as potentially uninhabitable and potentially habitable conditions for the Hadean Earth. Comprehensive climate models are required to better pinpoint the methane abundance cut-off for Hadean habitability, and are out of the scope of this work.

If conditions are too hot for ponds to exist, organic haze particles will settle on the dry surface and could potentially stockpile biomolecules until the atmosphere cools enough for ponds to form. However, it is unclear if any biomolecules in the solid haze particles can survive the hot surface conditions. Furthermore, we do not know how the biomolecule content of hazes scales with atmospheric methane abundance. Experiments suggest that HCN production in Hadean conditions is linearly dependent on atmospheric methane abundance \citep{Pearce_et_al2022b}, and HCN is a key precursor to many biomolecules including nucleobases and amino acids \citep{1961Natur.191.1193O,Reference437}. Therefore, we might also expect biomolecule production in organic hazes to scale linearly with atmospheric methane abundance. 

In this study, we analyze the nucleobase and amino acid content of solid organic haze particles produced in early Hadean atmospheric experiments containing high (5\%) and low (0.5\%) methane abundances. We also simulate the heating effects on the haze particles that settle on uninhabitable surfaces, and measure the biomolecule content of these particles to better understand the plausibility of biomolecule stockpiling. Lastly, we calculate the pond concentrations of nucleobases resulting from organic haze sources on the Hadean Earth using a numerical sources and sinks pond model. We compare these concentrations with the nucleobase pond concentrations resulting from {\it in situ} HCN chemistry, meteorites, and interplanetary dust from our previous studies \citep{Pearce_et_al2022a,Reference425}.

\section*{Methods}

\subsection*{Hadean Organic Haze Experiments}

Atmospheric simulations were performed with the Planetary Haze Research (PHAZER) experimental setup \citep{He_et_al2017}. PHAZER is a vacuum flow system and stainless steel chamber used to simulate chemistry and haze production in planetary atmospheres. Within the chamber is the option to attach different energy sources. For this work we attach two electrodes that allow us to produce a cold plasma discharge: a laboratory analog to short-wave upper atmospheric ultraviolet light ($\lesssim$ 110nm) \citep{Pearce_et_al2022b,Cable_et_al2012}. For a schematic of the PHAZER setup, see \citet{Pearce_et_al2022b}.

Our experimental protocol is similar to previous PHAZER studies \citep{Pearce_et_al2022b,Moran_et_al2020,He_et_al2020b,He_et_al2020a,He_et_al2019,Horst_et_al2018a,He_et_al2018b,He_et_al2018a}.
We start by cleaning the chamber and flanges using Alconox{\textsuperscript \textregistered} detergent powder and a scouring pad. We then use a fine grained (1500-grit) sandpaper to clean any remaining stuck-on solids from the chamber and flange walls, and the electrodes. The tight space between the flange and the electrodes is cleaned by scraping and twisting a wet paper towel into the area with a spatula repeatedly until the paper towel comes out clean. After a visual analysis shows no solids are present on the chamber walls, flange walls, and electrodes, these pieces are put in an ultrasonic cleaner for 3$\times$30 minutes. We then rinse the chamber, electrodes, and flanges with tap water, then HPLC-grade water, then methanol. Next, we put all the chamber pieces except the flange with the electrodes in the oven overnight at 80$^{\circ}$C. We visually inspect the chamber again to make sure no discoloration is present on the chamber pieces before assembling the chamber while wearing nitrile gloves. New copper gaskets are used to seal all flanges. Lastly, valves near the entry and exit points to the chamber are cleaned with methanol and cotton swabs.

We prepare the gas mixtures using high-purity gases (\ce{H2}-99.9999\%, \ce{N2}-99.9997\%, \ce{CO2}-99.999\%, and \ce{CH4}-99.999\%; Airgas). We vacuum and purge the lines with \ce{N2} multiple times before running our gas mixtures. We then turn on gas flow at 5 standard cubic centimeters per minute and initiate the 170 W/m$^2$ cold plasma source by applying a voltage differential of 6000 V between the electrodes. The cold plasma discharge produces electrons and ions roughly in the 5--15 eV range and are energetic enough to directly dissociate \ce{H2}, \ce{CO2}, \ce{CH4}, and \ce{N2}. The plasma doesn't significantly affect the neutral gas temperature, which remains approximately at room temperature. The exposure time of the gas to the plasma is approximately 2 seconds \citep{He_et_al2022a}.

Experiments are run for approximately six days until the mixing tanks are empty. The chamber pressure gauge varies from 2--3.5 Torr during that time. The readout variation is expected for \ce{H2}-dominant gas mixtures, as these pressure gauges are calibrated for direct readout of nitrogen or air. Gas flow and plasma are then turned off, the chamber valves are sealed off, and the chamber is transfered into a dry ($<$0.1 ppm \ce{H2O}), oxygen-free ($<$0.1 ppm \ce{O2}) \ce{N2} glovebox (I-lab 2GB; Inert Technology). Here, samples of solid organic haze particles are collected from the walls and base of the chamber with cleaned and heat-sterilized spatulas, and stored in small glass vials that were not previously used. Vial caps are screwed on and sealed with parafilm. No controls were produced at this stage.

Organic hazes are produced for two different Hadean atmospheric compositions: high methane content and low methane content. We collected 522 mg of solid haze particles from the high methane experiment, and 72 mg from the low methane experiment. The compositions of the gas mixtures are \ce{H2}:\ce{N2}:\ce{CO2}:\ce{CH4} = 89:5:1:5 and 93.5:5:1:0.5 for the high and low methane experiments, respectively.

Given the potency of methane as a greenhouse gas, the high (5\%) methane atmosphere may have had surface temperatures above the boiling point of water. Therefore, as an initial study of the effects of high temperatures on our samples, we take two portions of the organic haze particles collected from the high methane experiment and heat them at 200 $^{\circ}$C in an oven for 1 day and 7 days, respectively. To do this without exposing the sample to oxygen, we place the glass vial (with cap off) containing the organic haze particles in a Fisherbrand heavy glass desiccator while in the \ce{N2} glovebox and lubricate the contact between the dessicator base and lid with high-vacuum silicon grease (rated to 204 $^{\circ}$C) to make an air-tight seal. We then transfer the desiccator to the oven for heating. Once heating is complete, we transfer the dessicator back to the glovebox to remove the sample and store it for later biomolecule analysis.

\subsection*{Biomolecule Analysis}

Biomolecule analyses are performed on an Agilent 7000c triple quadrupole gas chromatograph/mass spectrometer (GC/MS) using a highly selective and sensitive GC/MS/MS protocol developed for detecting nucleobases and amino acids in organic haze particles \citep{Sebree_et_al2018}. GC/MS/MS differs from regular GC/MS in that the first quadrupole traps only the selected parent masses within each chosen range of retention times. Then, the ions are transferred to the second quadrupole where they are fragmented by collision with neutral gas. Finally, the fragment ions are passed along to the third quadrupole, where only two masses associated with the fragmentation (quantifier and qualifier) are allowed to pass through the detector. The benefit of this technique is the increased signal-to-noise that arises from 1) the preconcentration of the species of interest at the first quadrupole, and 2) the supression of other species with the same GC retention time as the species of interest (as they are ignored at the first quadrupole), and 3) the further supression of noise by only allowing two daughter masses from the intended parent mass to pass through the detector. This method differs from an extracted ion chromatogram, for which quantitation ions are extracted from a total ion chromatogram (TIC) containing all masses. The GC/MS/MS protocol instead filters out all masses that we are not looking for (in each gate) prior to detection using the three quadrupoles. In the case where there are two species in the same gate with the same retention times, e.g., uracil and proline (gate 10), we calculate the relative peak percentages using the relative intensities of the quantifier peaks in the mass spectra (see Figure~\ref{Gate10Peaks}). This procedure is best suited to analyze complex samples like ours that require higher sensitivity. The retention times for the GC/MS/MS gates were found and developed in \citet{Sebree_et_al2018} by running amino acid and nucleobase standards through regular GC/MS and analyzing at the mass spectra for each GC peak. We refer the reader to \citet{Sebree_et_al2018} for more information about that procedure.

We first carry out GC/MS/MS validation, limit of detection, and calibration curve development on two sets of standards: 1) A physiological amino acid standard (Sigma Aldrich) with 27 detectable components at 2.5 $\mu$mol mL$^{-1}$ biomolecule$^{-1}$, and 2) a nucleobase standard (Sigma Aldrich, all 99.0\%) containing seven nucleobases. The seven nucleobases are dissolved in a 0.1M solution of NaOH at a concentration of 0.5 mg mL$^{-1}$ nucleobase$^{-1}$. To find limits of detection and quantification, and produce calibration curves, dilutions of $\frac{5}{100000}$, $\frac{20}{100000}$, and $\frac{200}{100000}$ are made of each standard to provide a range of base concentrations to sample different volumes from for GC/MS/MS analysis.

The standard sample preparation procedure is as follows: 1) multiple volumes in the range of 5--50 $\mu$L of each base standard solution are added to separate GC vials and dried in the oven at 40 $^{\circ}$C under \ce{N2} flow. After all visible liquid evaporated, vials are washed with 100 $\mu$L of \ce{CH2Cl2} and dried again at 40 $^{\circ}$C under \ce{N2} flow. Next, 30 $\mu$L of dimethylformamide (DMF) and 30 $\mu$L of N-tert-butyldimethylsilyl-N-methyltrifluoroacetamide (MTBSTFA) are added to the dried GC vials and the solutions are left to derivatize for 30 minutes under \ce{N2} flow at 80 $^{\circ}$C. Finally, 100 $\mu$L of \ce{CH2Cl2} is added to the solutions, and 0.5, 0.8, 1, or 2 $\mu$L of the solution are injected into the GC triple quad running in MS/MS mode for analysis. We do multiple injection volumes in order to get several data points for our calibration curves for a single sample concentration. Each sample takes approximately 25 minutes to run.

Separation of compounds is performed using a Restek capillary column (RTX-5MS) with the helium flow held constant at 1.3 mL min$^{-1}$. The operating temperature is 100 $^{\circ}$C, ramping up at a rate of 10 $^{\circ}$C min$^{-1}$ to a final temperature of 270 $^{\circ}$C; held for 11.5 minutes. Collisionally induced dissociation energies of 10--50 eV are used during MS/MS operations in the MRM scan mode.

In Figure~\ref{NucleobaseStandard}, we display the GC/MS/MS chromatograms for nine nucleobase standard concentrations, and in Figure~\ref{PhysioStandard}, we display the GC/MS/MS chromatograms for nine physiological amino acid standard concentrations. We ran the nucleobase standards first, and it became apparent once running some subsequent tests with methanol and MTBSTFA derivatizer that underivatized nucleobases from the high concentration runs stick around in the GC inlet. This is to be expected, given that a) no derivatization protocol yields 100\% derivatized biomolecules, and b) MTBSTFA derivatization is a reversible reaction. This hypothesis is validated by two sets of tests. First, we find that nucleobase peaks do not show up in the methanol runs, but do show up in the derivatizer runs (see Figure~\ref{MeOH_spectra}). Second, we test a washing protocol for the GC inlet, whereby we track the decrease in the nucleobase peaks after subsequent GC/MS/MS runs with a MTBSTFA derivatizer. If there are underivatized nucleobases leftover in the GC inlet, the MTBSTFA would derivatize them, allow them to vaporize in the inlet and enter the GC column. This would result in a reduction of the nucleobase peaks with each wash run. In Figure~\ref{Hypoxanthine_Wash}, we display the results for the hypoxanthine peak across 17 wash runs. We see that the hypoxanthine peak decreases with every subsequent wash.

Following this discovery, we initiate a new protocol to wash the GC inlet twice with a procedural blank containing MTBSTFA before any individual GC/MS/MS run on organic haze particles. We subtract the second blank chromatogram from the haze particle chromatogram for analysis. We find that there is no retention time difference between haze particle and blank runs. The linearity of our calibration curves suggests that our protocol can be used for quantification, as inconsistent levels of MTBSTFA derivitization or degradation would show up as non-linearities in the calibration curves.

Peak areas are integrated using the average noise floor for each GC gate (see Table~\ref{RetentionTimes} for list of GC gates). Uncertainties are calculated by integrating the peak areas with noise floor adjusted by $\pm$ 1$\sigma$. We wrote a Python code to automatically perform these calculations for pre-selected GC gates and noise regions (see \citet{Pearce_Chromatogram_2023}). The code outputs an interactive visualization of the integrated peaks, allowing the user to validate the retention times and signal-to-noise ratios. Peaks falling below an SNR of 3 are screened out in this process. See Figure~\ref{Peak_Area_Example} for a screenshot of the Python code and an example peak area calculation.

\begin{figure*}[tp]
\centering
\includegraphics[width=\linewidth]{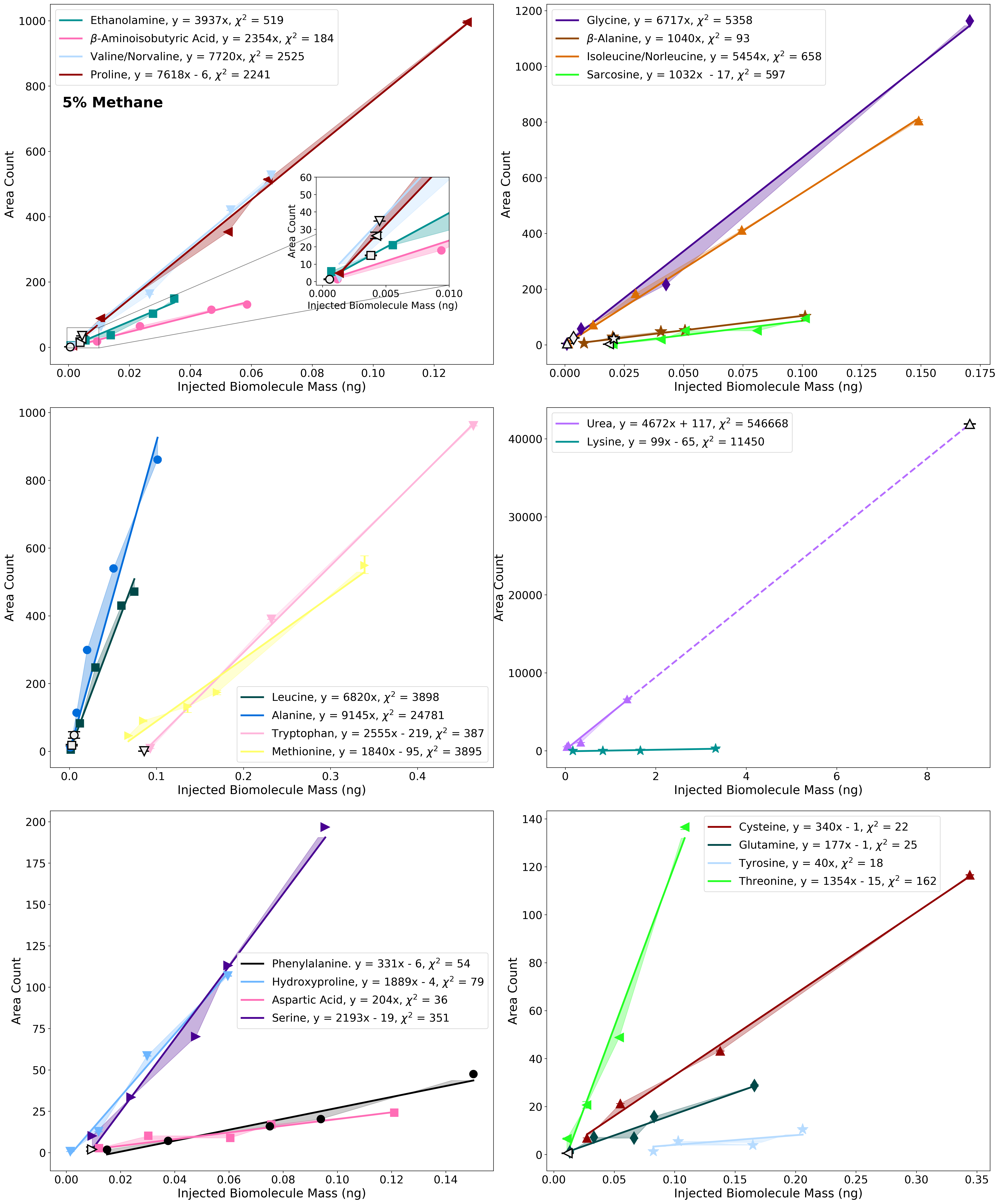}
\caption{Calibration curves calculated for 22 amino acids and other organics. Equations of the lines of best fit and $\chi^2$ values are added to the legend. Uncertainties are the shaded regions calculated as the variance from a model linearly connecting each data point. Peak areas for the 5\% \ce{CH4} haze particles are added to the plots to demonstrate our quantification calculations. \label{PAmino_Cal_CurveH5}}
\end{figure*}

In Figures~\ref{Nucleobase_Cal_Curve} and \ref{PAmino_Cal_CurveH5}, we display the calculated calibration curves for the nucleobases and amino acids/other biomolecules, respectively. Calibration curves allow us to quantify the concentrations of each of these species in our organic haze particles. Calibration curves are calculated by plotting the integrated GC peak areas from our standards for a range of injected masses for each species, and performing a least squares linear fit to the data. Injected masses are calculated using the concentration of each species in the standard, the volume of the standard put into the GC vial, the volume by which it was diluted during derivatization/preparation, and the volume injected from the GC vial into the GC inlet. We select 3--5 data points for the linear fits that are within the range of peak areas found in our haze particles, and whose fit has the lowest $\chi^2$ value. Urea is the only species for which we did not use the lowest detected injection mass from our standards (see Table~
\ref{SNRTable} for details on SNR of lowest detected injection masses). Uncertainties are included on the calibration curves based on their variance from a model linearly connecting each data point. 

Organic haze particles are prepared for analysis similarly to \citet{Sebree_et_al2018}. First, the organic haze particles are weighed and and dissolved in a 50:50 solution of methanol and acetone at 5.0 mg mL$^{-1}$. The vials are then shaken and centrifuged at 10,000 rpm for 10 minutes, after which they are rotated 180 degrees and centrifuged for another 10 minutes. The supernatants are separated from the solids with a Pasteur pipette and transferred to storage vials. We dry and derivatize 50 $\mu$L of the storage solutions using the same methods described above. 1 $\mu$L of the solutions are injected into the GC triple quad running in MS/MS mode for analysis. Two blank solutions of 50:50 methanol/acetone are also carried through the same procedure as a contamination check and to provide the dervatizer to wash the GC inlet as described in the washing protocol above.

\subsection*{Warm Little Pond Models}

To calculate the nucleobase concentrations in warm little ponds from organic hazes, we use a comprehensive sources and sinks numerical pond model. The model was originally developed in \citet{Reference425} to compare meteorites and interplanetary dust as sources of nucleobases to ponds, and was modified in \citet{Pearce_et_al2022a} to include {\it in situ} production from atmospherically sourced HCN. The model is based on ordinary differetial equations which are solved numerically and correspond to the sources and sinks for both water and nucleobases. Water in this model comes exclusively from precipitation, and leaves via evaporation and seepage (through pores in the base of the pond). The precipitation rates we use match the ``Intermediate Environment'' from \citet{Reference425}, which produce naturally occurring seasonal wet-dry cycles. Nucleobases in this model are sourced from organic hazes, and are removed via photodissociation, hydrolysis, and seepage. We model a 1-meter radius and depth cylindrical pond at 65 $^{\circ}$C to represent a typical pond on the hot Hadean Earth. See \citet{Reference425} and \citet{Pearce_et_al2022a} for details on rate equations, sources of common rate data, and model schema. The numerical pond model is written in Python and a release made available for download \citep{Pearce_Pond_2023}.

In this work, we add nucleobase deposition from organic hazes to the sources and sinks pond model. The rate for this deposition is calculated using the following equation,

\begin{equation}
\dot{m_N} = [N]\dot{m}_H
\end{equation}
where $[N]$ is the concentration of nucleobase $N$ within the organic haze particles (grams of nucleobase / grams of haze particles), and $\dot{m}_H$ is the mass deposition of haze particles on the Hadean Earth (grams / year).

We use the mass flux proportionality equation empirically derived by \citet{Trainer_et_al2006} to estimate the minimum and maximum haze particle deposition rates for the Hadean Earth at 4.4 billion years ago (bya). The equation is.

\begin{equation}
\dot{m}_H = \beta F_T \gamma A_{\bigoplus} \left(\frac{I_{0,4.4}}{I_{0,T}}\right)^n \left(\frac{\chi_{c,4.4}}{\chi_{c,T}}\right)
\end{equation}
where $\beta$ is the empirically derived reduction factor of 0.45 from having \ce{H2}-dominant conditions and both \ce{CH4} and \ce{CO2} in similar abundances the atmosphere \citep{DeWitt_et_al2009}, $F_T$ is the haze particle flux deposited on Titan's surface (g cm$^{-2}$ s$^{-1}$), $\gamma$ is the number of seconds in a year, $A_{\bigoplus}$ is the area of the Earth's surface (cm$^{2}$), $I_{0,4.4}$ is the solar flux on Earth at 4.4 bya, $I_{0,T}$ is the solar flux on Titan, $n$ = 1 or 2 for the two photochemical pathways of haze production (A or B) derived in \citet{Trainer_et_al2006}, $\chi_{c,4.4}$ is the atmospheric mixing ratio of \ce{CH4} on Earth at 4.4 bya, and $\chi_{c,T}$ is the atmospheric mixing ratio of \ce{CH4} on Titan.

For our calculation, we employ the same values as used in \citet{Trainer_et_al2006}, except two parameters. 1) for $I_{0,4.4}$, we multiply the present day solar flux on Earth by a factor of 20 to account for the at minimum 20 times stronger emission in the extreme ultraviolet range (1--120 nm) at 4.4 bya \citep{Ribas_et_al2005}. 2) We use 0.5\% \ce{CH4} for $\chi_{c,4.4}$, to match our low methane, potentially habitable Hadean experiment.

Considering exponential $n$ values of 1 or 2, we obtain a minimum haze particle mass deposition rate of 4.9$\times$10$^{14}$ g yr$^{-1}$ and a maximum haze particle mass deposition rate of 8.9$\times$10$^{17}$ g yr$^{-1}$. These values are approximately two orders of magnitude higher than the organic haze production rates estimated by \citet{Trainer_et_al2006} for the Archean Earth.

\section*{Results}

\subsection*{Biomolecules in Organic Hazes}

\begin{figure*}[!hbtp]
\centering
\includegraphics[width=\linewidth]{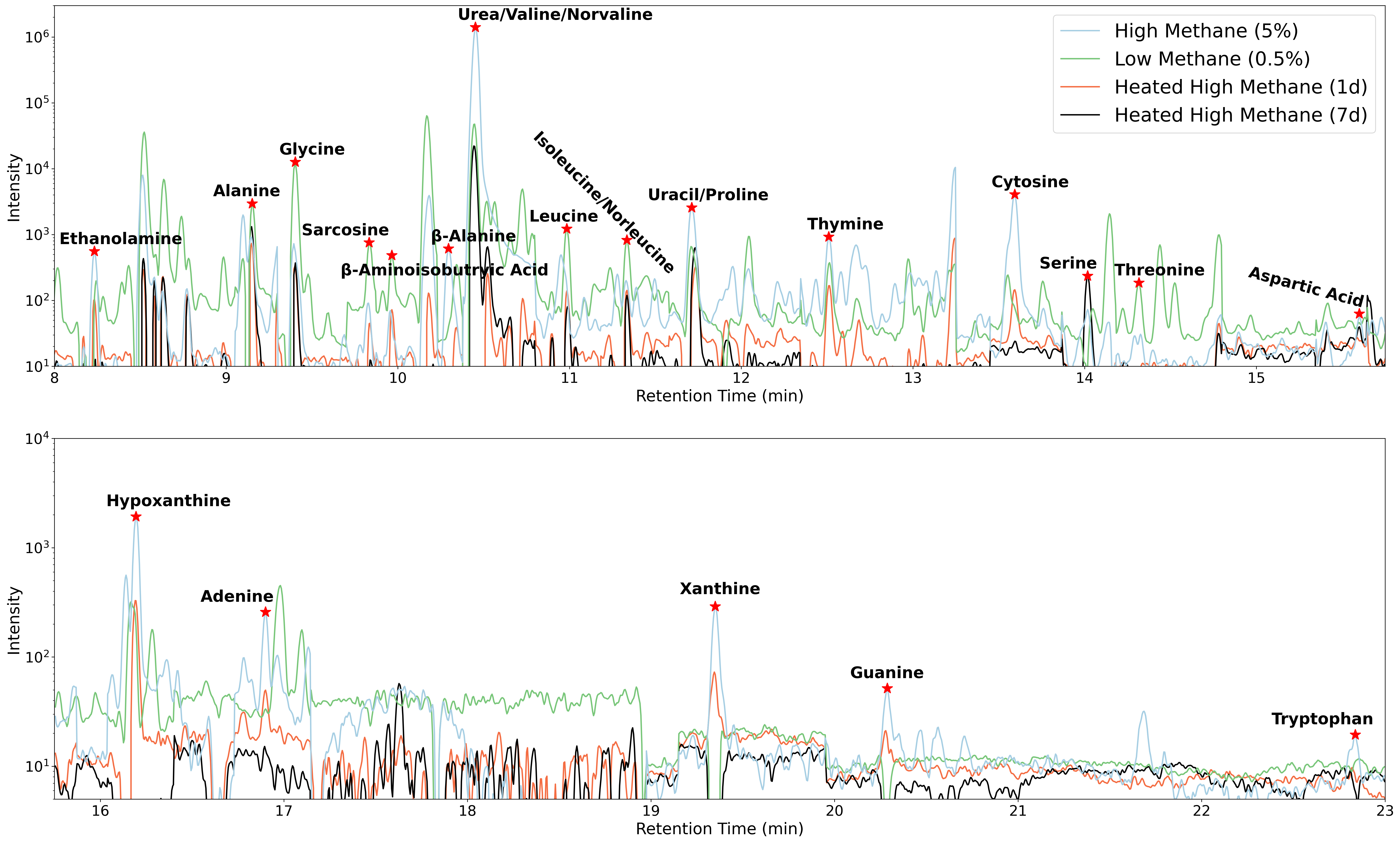}
\caption{Blank-subtracted GC/MS/MS chromatograms for extracted samples of the high methane organic haze particles, the low methane organic haze particles, and the one- and seven-day heated high methane organic haze particles. \label{HazeSpectra}}
\end{figure*}

In Figure~\ref{HazeSpectra}, we display the GC/MS/MS chromatograms for the Hadean haze particle samples from the high and low methane experiments, as well as the high methane Hadean haze particle samples heated to 200 $^{\circ}$C for 1 and 7 days. Seven nucleobases, nine proteinogenic amino acids, and five other organics (including some non-proteinogenic amino acids) were detected in total. We summarize the detected biomolecules and their concentrations in Table~\ref{ConcentrationTable}.



\begin{table*}[ht!]
\centering
\caption{Measured mass fractions of nucleobases, amino acids, and other organics in four Hadean Earth experiments. Uncertainties are the combination of GC peak area uncertainties, and the calibration curve uncertainties (see Appendix for explanation). In some cases, only an upper limit is provided for measurements that fall above the limit of detection, but below the limit of quantification (see Table~
\ref{LODLOQTable} for details). Units are $\mu$g biomolecule / g haze particles, or ppm. \label{ConcentrationTable}} 
\begin{tabular}{lcccc}
\\
\multicolumn{1}{l}{Biomolecule} &  
\multicolumn{1}{l}{High Methane} & 
\multicolumn{1}{l}{Heated High Methane (1d)} & 
\multicolumn{1}{l}{Heated High Methane (7d)} & 
\multicolumn{1}{l}{Low Methane}
\\[+2mm] \hline \\[-2mm]
{\bf Nucleobases} & & & \\
Xanthine & 62$\substack{+3 \\ -1}$ & 17.8$\substack{+5.3 \\ -0.1}$ & - &  - \\
Hypoxanthine & 47$\substack{+2 \\ -1}$ & 18.5$\substack{+0.2 \\ -0.3}$ & - & 20.8$\substack{+0.6 \\ -0.0}$ \\
Cytosine & 19.2$\substack{+0.5 \\ -0.3}$ & 2.9$\substack{+0.1 \\ -0.9}$ & - & 3.3$\substack{+0.1 \\ -0.6}$ \\
Guanine & 14.8$\substack{+2.7 \\ -1.6}$ & 10.8$\substack{+0.3 \\ -4.5}$ & - & - \\
Uracil & 5.8$\substack{+0.7 \\ -0.2}$ & 3.0$\substack{+0.0 \\ -1.5}$ & 2.8$\substack{+0.0 \\ -1.5}$ & 3.8$\substack{+0.1 \\ -1.7}$ \\
Adenine & 2.1$\substack{+0.5 \\ -0.2}$ & 1.58$\substack{+0.04 \\ -0.14}$ & - & 4.7$\substack{+1.6 \\ -0.1}$ \\
Thymine & 1.8$\substack{+0.1 \\ -0.5}$ & 1.6$\substack{+0.0 \\ -0.1}$ & - & 1.7$\substack{+0.0 \\ -0.3}$ \\
{\bf Proteinogenic Amino Acids} & & & \\
Tryptophan & $<$50$\substack{+4 \\ -0}$ & - & - & -\\
Threonine & 6.7$\substack{+0.1 \\ -0.4}$ & - & - & 8.8$\substack{+1.9 \\ -1.9}$\\
Serine & 5.5$\substack{+0.4 \\ -0.3}$ & - & 6.8$\substack{+0.1 \\ -1.3}$ & 6.0$\substack{+0.2 \\ -0.2}$\\
Alanine & 3.1$\substack{+0.6 \\ -1.4}$ & 1.1$\substack{+0.0 \\ -0.4}$ & 2.0$\substack{+0.0 \\ -0.7}$ & 5.0$\substack{+0.1 \\ -1.8}$\\
Valine/Norvaline & $<$2.6$\substack{+1.2 \\ -0.0}$ & 0.4$\substack{+0.9 \\ -0.1}$ & 1.5$\substack{+1.0 \\ -0.1}$ & $<$4.1$\substack{+8.0 \\ -3.3}$\\
Proline & 2.5$\substack{+0.1 \\ -0.4}$  & 1.09$\substack{+0.06 \\ -0.04}$ & 1.72$\substack{+0.03 \\ -0.02}$ & 1.34$\substack{+0.04 \\ -0.20}$ \\
Glycine & 2.1$\substack{+0.1 \\ -0.3}$ & 0.7$\substack{+0.2 \\ -0.0}$  & 0.63$\substack{+0.08 \\ -0.01}$ & 33$\substack{+6 \\ -1}$\\
Isoleucine/Norleucine & $<$0.54$\substack{+0.03 \\ -0.03}$ & $<$0.32$\substack{+0.03 \\ -0.02}$ & 0.23$\substack{+0.02 \\ -0.02}$ & 2.1$\substack{+0.4 \\ -0.2}$\\
Leucine & 1.5$\substack{+0.5 \\ -0.2}$ & 0.24$\substack{+0.01 \\ -0.01}$ & 0.15$\substack{+0.02 \\ -0.01}$ & 2.8$\substack{+0.2 \\ -0.2}$\\
{\bf Other Biomolecules} & & & \\
Urea & 5204$^a\substack{+2 \\ -2}$ & - & 65$\substack{+0 \\ -32}$ & 170$\substack{+13 \\ -10}$\\
$\beta$-Alanine & 12.0$\substack{+0.1 \\ -2.4}$ &  $<$0.4$\substack{+4.8 \\ -0.1}$ & - & 7.4$\substack{+0.1 \\ -0.1}$ \\
Sarcosine & $<$10.8$\substack{+0.2 \\ -0.1}$ & $<$11.1$\substack{+1.9 \\ -0.2}$ & - & 23$\substack{+3 \\ -0}$ \\
Ethanolamine & 2.2$\substack{+0.0 \\ -0.1}$ & $<$0.27$\substack{+0.18 \\ -0.01}$ & - & - \\
$\beta$-Aminoisobutyric Acid & $<$0.33$\substack{+0.40 \\ -0.04}$ & $<$0.33$\substack{+0.42 \\ -0.03}$ & - & $<$2.6$\substack{+1.2 \\ -0.3}$ \\
\\[-2mm] \hline
\multicolumn{5}{l}{\footnotesize $^a$ based on upper bound extrapolation of calibration curve} \\
\end{tabular}
\end{table*}

\begin{figure}[!hbtp]
\centering
\includegraphics[width=\linewidth]{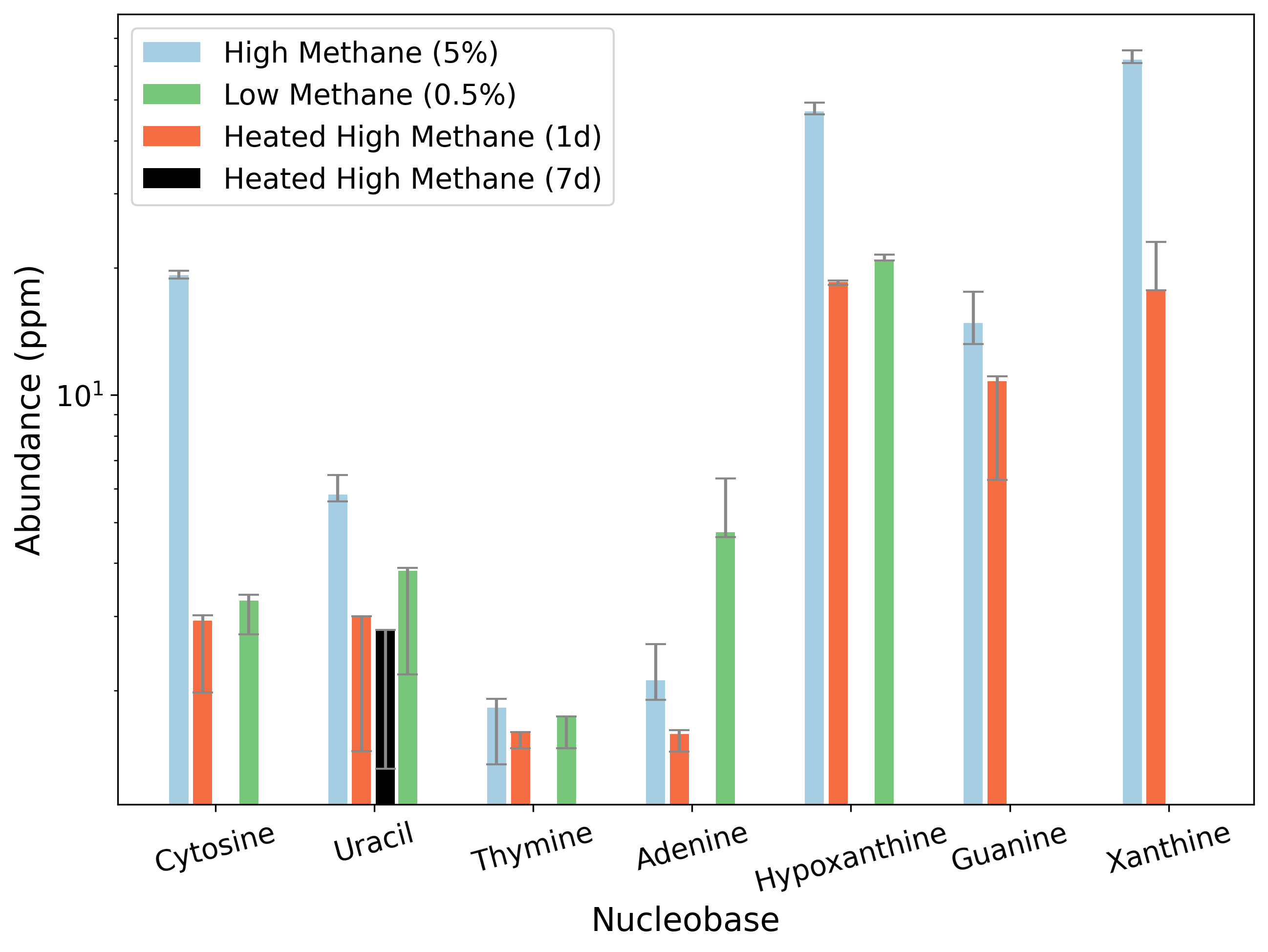}
\caption{Nucleobase abundances measured in the four organic haze particle samples in this study. Nucleobase order is from lowest to highest molar mass. Units are $\mu$g biomolecule / g haze particles, or ppm. \label{Nucleobases}}
\end{figure}

In Figure~\ref{Nucleobases}, we display the nucleobase concentrations in our four Hadean organic haze particle samples. All seven nucleobases are detected in the high methane organic haze particles, as well as the haze particles heated for 1 day. Guanine and xanthine are not detected in the low methane organic haze particles, and only uracil is detected in the haze particles heated for 7 days. Concentrations are highest in the high methane experiment, with the exception of adenine, for which the low methane haze particles have the highest concentration. The nucleobases with the highest concentrations in the high methane experiment haze particles are xanthine, hypoxanthine, cytosine, and guanine, ranging from $\sim$15--62 parts-per-million (ppm). Concentrations are lowest for thymine, adenine, and uracil, ranging from $\sim$2--6 ppm for this experiment. The low methane experimental haze particles range from being 6 times lower to 2 times higher in nucelobase concentrations than the high methane experimental haze particles. In total, the nucleobase content of the high methane haze particles is 4.5 times higher than the nucleobase content of the low methane haze particles. The high methane haze particles heated for 1 day have 1--7 times lower nucleobase concentrations than the unheated high methane haze particles. Overall, heating the high methane haze particles for a day leads to a factor of three reduction in total nucleobase content. All nucleobases are destroyed after 7 days of heating, except uracil which has the same concentration in the samples heated for 1 and 7 days. This may suggest uracil reaches a steady state to thermal decay at 200 $^{\circ}$C after only 1 day.


\begin{figure*}[!hbtp]
\centering
\includegraphics[width=\linewidth]{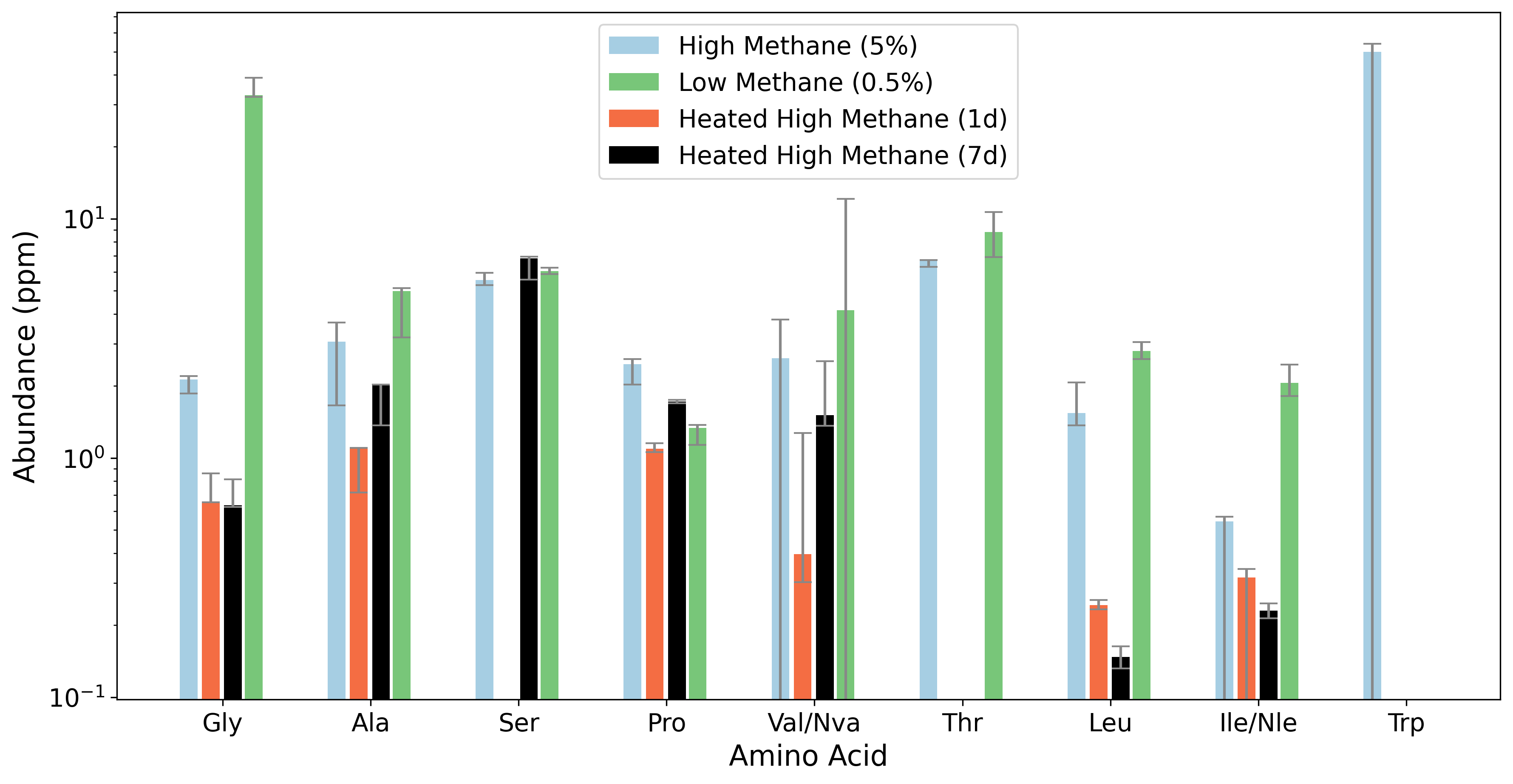}
\caption{Proteinogenic amino acid abundances measured in the four organic haze particle samples in this study. Amino acid order is from lowest to highest molar mass. Units are $\mu$g biomolecule / g haze particles, or ppm. \label{Aminos}}
\end{figure*}

In Figure~\ref{Aminos}, we display the proteinogenic amino acid concentrations in our four organic haze particle samples. Differences in individual proteinogenic amino acid concentrations between the high methane experimental haze particles and the low methane experimental haze particles vary by a factor of $\sim$1--16. However, the total amino acid content of the high methane haze particles is only a factor of 1.2 greater than the amino acid content of the low methane haze particles. Tryptophan is only detected in the high methane haze samples. Proline is higher in concentration in the high methane experimental haze particles, whereas glycine, leucine, and isoleucine/norleucine are higher in concentration in the low methane experimental haze particles. Alanine, serine, valine/norvaline, and threonine have similar concentrations in both haze particle samples.

Heating the high methane haze particles for 1 day destroyed a few amino acids, including serine, threonine, and tryptophan, and reduced the concentrations of glycine, alanine, proline, valine/norvaline, and leucine by factors of 1--13. Interestingly, heating the high methane haze particles for 7 days only reduced the concentrations of leucine and isoleucine/norleucine from their 1 day heating concentrations. Alanine, proline, and valine/norvaline increased in concentration after the additional heating, and serine was detected after not being present in the 1 day heated haze particles. This may suggest that alanine, proline, valine/norvaline, and serine are thermal decay products of the organic haze particles.

\begin{figure}[!hbtp]
\centering
\includegraphics[width=\linewidth]{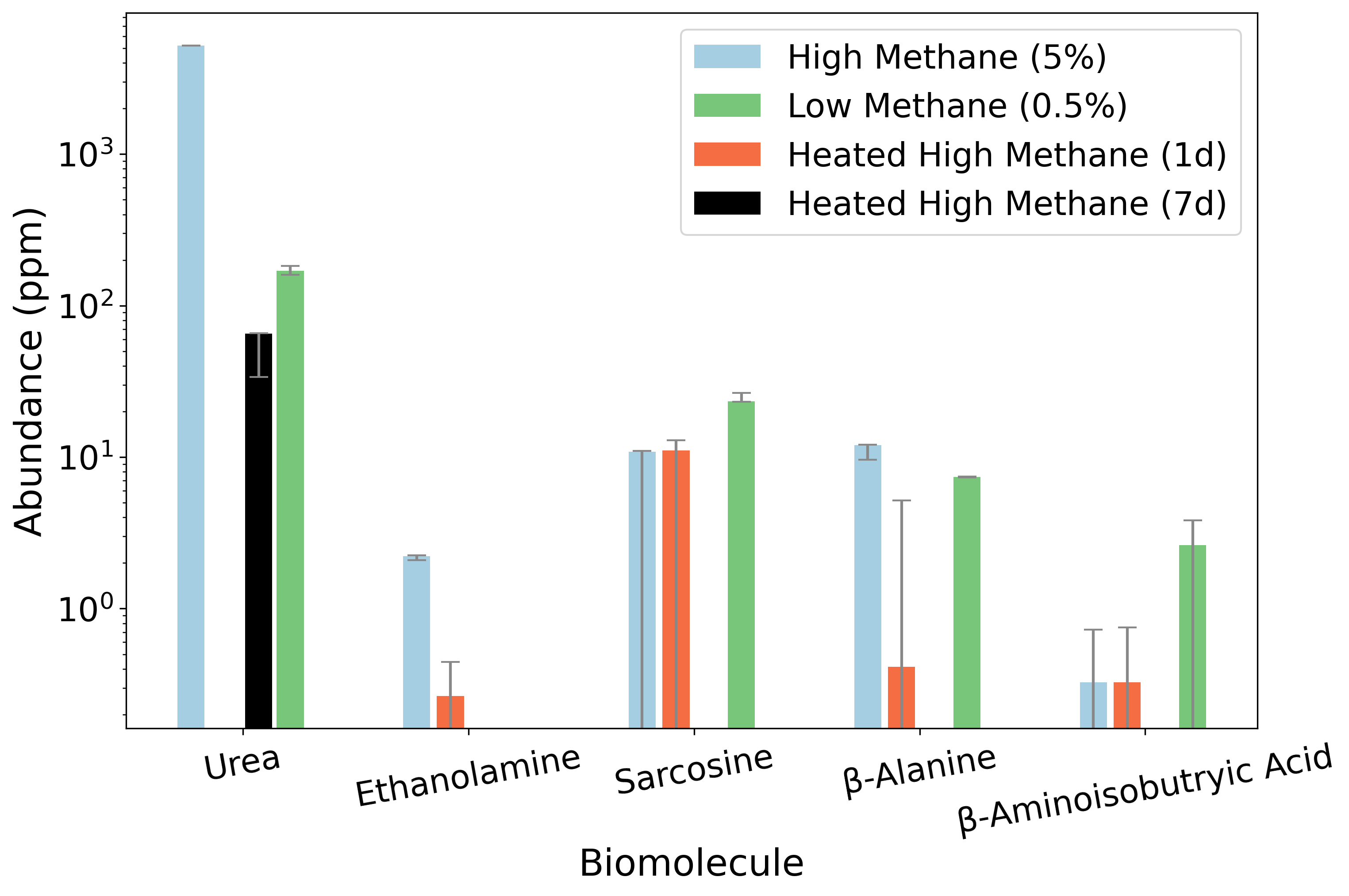}
\caption{Non-proteinogenic amino acid and other organic abundances measured in the four organic haze particle samples in this study. Biomolecule order is from lowest to highest molar mass. Units are $\mu$g biomolecule / g haze particles, or ppm. \label{NonProtAminos}}
\end{figure}

Finally, in Figure~\ref{NonProtAminos}, we display the non-proteinogenic amino acids and other biomolecule concentrations in our photochemically produced organic haze particles. Urea, ethanolamine, and $\beta$-alanine concentrations are highest in the high methane organic haze particles, and sarcosine and $\beta$-aminoisobutyric acid concentrations are highest in the low methane organic haze particles. After 1 day of heating the high methane haze particles, urea is no longer detected, and the other four biomolecules either remain the same concentration as the unheated haze particles, or decrease by a factor of 8--30. Interestingly, after 7 days of heating, the only detectable biomolecule remaining is urea, which was undetected after 1 day of heating. This may suggest that urea is a longer-term decomposition product of other biomolecules in the organic haze particles.

\subsection*{Biomolecule Concentrations in Warm Little Ponds}

\begin{figure*}[!hbtp]
\centering
\includegraphics[width=\linewidth]{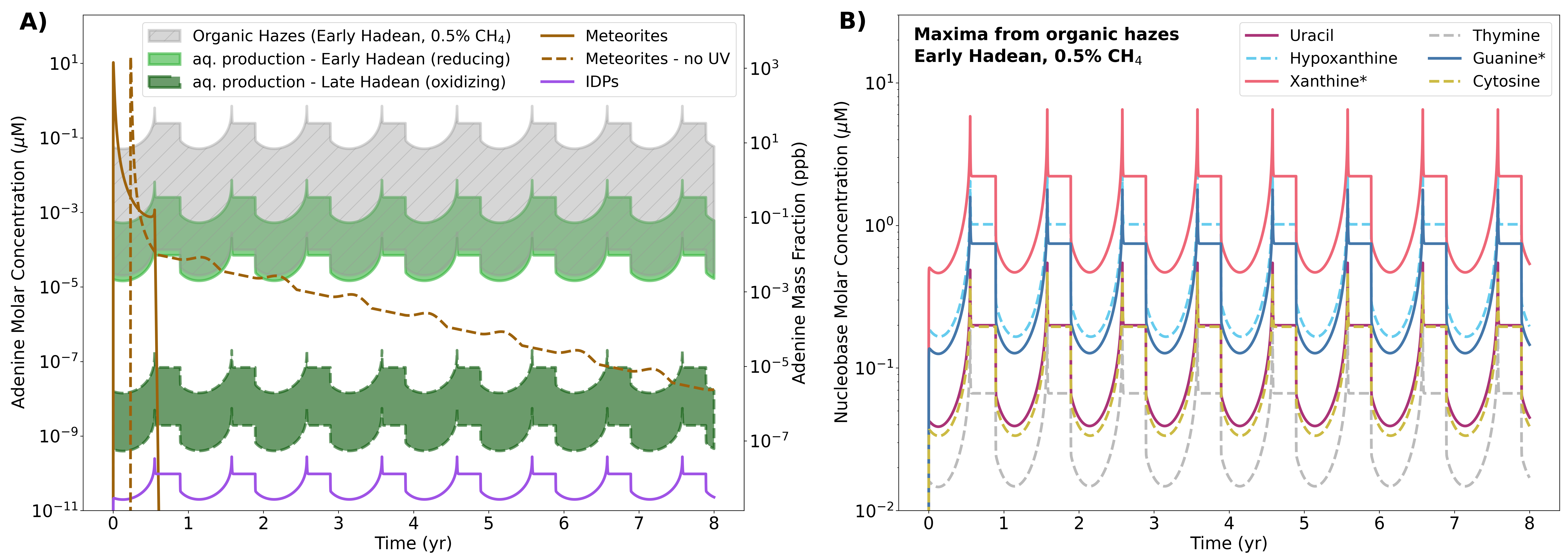}
\caption{A) Model pond concentrations of adenine from organic haze particle deposition at 4.4 billion years ago (bya) for a 0.5\% methane atmosphere (this work) compared with meteoritic delivery \citep{Reference425}, interplanetary dust particle (IDP) delivery \citep{Reference425}, and {\it in situ} production from atmospherically sourced HCN for two different periods during the Hadean \citep{Pearce_et_al2022a}. Concentrations are calculated using a comprehensive sources and sinks pond model originally developed in \citet{Reference425} that has a $\sim$6 month wet and $\sim$6 month dry cycle; the one exception is the ``Meteorites - no UV'' model, which never dries up and for which UV is never turned on due to water shielding. B) Maximum pond concentrations of 6 other nucleobases from organic haze particle deposition at 4.4 bya (this work). *These values are calculated using the high methane (5\%) \ce{CH4} experimental haze particle concentration, as xanthine and guanine were not detected in the low methane (0.5\%) organic haze particles (see Table~\ref{ConcentrationTable}). \label{WarmLittlePondModel}}
\end{figure*}

In Figure~\ref{WarmLittlePondModel}A, we display the concentration of adenine modeled in a cylindrical warm little pond with a 1-meter radius and depth over an 8 year period, from five different sources. The adenine concentrations from organic hazes were modeled in this work and are based on the potentially habitable low methane (0.5\%) experimental haze particles. The adenine concentrations from meteoritic and interplanetary dust particle delivery were modeled in \citet{Reference425}, and the concentrations from {\it in situ} production from atmospherically sourced HCN were modeled in \citet{Pearce_et_al2022a}. The concentrations from steady sources cycle with a 1 year period due to the seasonal wet-dry cycles produced from evaporation, seepage, and precipitation.  The flat top pattern in the concentrations are produced in the dry phase when the adenine influx and photodestruction reach a steady state. Meteorites are single deposition events, and it is improbable that a single pond would have had more than one meteorite deposition \citep{Reference425}; therefore, adenine concentrations from these sources are fleeting.

The range of adenine concentrations from organic hazes are based on the minimum and maximum haze particle deposition rates of 4.9$\times$10$^{14}$ and 8.9$\times$10$^{17}$ g yr$^{-1}$ (see Methods), respectively, as well as the uncertainty in the adenine concentration within organic haze particles, as measured in this study (see Table~\ref{ConcentrationTable}).

Interestingly, organic hazes potentially delivered the highest stable adenine concentrations to ponds at 4.4 bya of up to $\sim$0.7 $\mu$M ($\sim$100 ppb). However, we note that the adenine concentrations from organic hazes during the early Hadean (4.4 bya) agree within uncertainty with the adenine concentrations produced within the pond from atmospherically sourced HCN at the same period. This suggests that both of these sources may have played a part in the origin of adenine in warm little ponds. The pattern of adenine concentration over the year is due to pond level variation (i.e., evaporation and precipitation), as well as the adenine sinks within the pond, i.e., photodissociation, hydrolysis, and seepage through pores at the base. Peak adenine pond concentrations from meteorites are one order of magnitude higher ($\sim$10 $\mu$M) than the maximum concentrations from organic hazes; however, meteoritically delivered adenine depletes within weeks, whereas organic hazes are a constant feeding source. As solar activity calms throughout the Hadean, haze production will decrease slightly. Ultraviolet intensity in the 92--120 nm range decreases from 20 times present day to 10 times present day from 4.4 to 4.1 bya. Considering this, we expect at least a factor of 4 decrease in the maximum adenine concentrations in ponds from organic hazes over a 300 million year period during the Hadean.

In Figure~\ref{WarmLittlePondModel}B, we display the maximum modeled pond concentrations of the other six nucleobases in this study sourced from organic haze at 4.4 bya. Eight years of integration are shown; however, a steady solution is achieved in just two years due to the balance of sources and sinks. The nucleobases that have the highest pond concentrations are xanthine, hypoxanthine, and guanine at $\sim$1.8--6.5 $\mu$M. We note, however, that the xanthine and guanine concentrations for this plot are based on the high methane (5\%) experimental haze particles as they were undetected in the low methane (0.5\%) experimental haze particles. Maximum pond concentrations of uracil, cytosine, and thymine are $\sim$0.21--0.55 $\mu$M. Differences in pond concentrations are due mainly to the differences in nucleobase abundances measured in our experimental haze particle samples (see Table~\ref{ConcentrationTable}).

\section*{Discussion}

The nucleobase and amino acid contents of organic haze particles decrease by a factor of 4.5, and 1.2, respectively, when lowering the atmospheric methane in our experiments by an order of magnitude. This may not be surprising, given there are a factor of 4 more carbon atoms in the high methane atmospheric experiment, and nucleobases and amino acids are based on carbon backbones. However, haze production also becomes less efficient when reducing methane in the simulated atmosphere, as the total mass of solid haze particles collected decreased by a factor of seven. Considering these two aspects, there is an approximate linearity between nucleobase and other biomolecule influx from organic hazes, and atmospheric methane abundance. This suggests that the ideal conditions during the Hadean eon for the origin of life would be when methane content is high enough to produce lots of organic haze, but is not too high to make the surface uninhabitable. Organic stockpiling does not seem to be a viable option for the origin of life's building blocks, as in just 24 hours of heating at 200 $^{\circ}$C, total nucleobase concentrations in organic haze particles decrease by a factor of 3, and after 7 days of heating at at 200 $^{\circ}$C, only uracil survives. Furthermore, heating the organic haze particle samples completely destroys some key proteinogenic amino acids such as threonine, and tryptophan.

Organic hazes as a source of life's building blocks is a compatible hypothesis with a big (400--500 km) iron-containing impactor scenario for the origin of life (e.g., \citet{Wogan_et_al2023}. The big impact produces the \ce{H2} and provides the heat ($>$950 K) needed to thermodynamically produce lots of methane - which we find to be linearly proportional to biomolecule influx from organic hazes. In the absence of a large impactor, models presently suggest that terrestrial sources such as hydrothermal environments may not produce enough atmospheric methane to build up to levels near 0.5\% which is ideal for abundant organic haze particle production \citep{Guzman-Marmolejo2013,Miyazaki_Korenaga_2022}. In \citet{Wogan_et_al2023}, HCN and HCCCN fluxes to ponds are calculated as sources of biomolecule precursors; however, we learn from this work that organic hazes would also be abundantly produced in such a methane-rich atmosphere, and would provide RNA building blocks to ponds in $\mu$M concentrations.

As far as we know, the origin of RNA on the Hadean Earth requires a reasonably high, stable concentration of RNA building blocks in pond environments. The exact concentration required is somewhat uncertain; however, a reasonable lower bound might be $\sim$100 $\mu$M. This is the lowest concentration that experiments have used to react nucleobases with ribose and phosphate to produce nucleotides \citep{1963Natur.199..222P}. Our pond models suggest that organic hazes could deliver up to $\sim$0.2--6.5 $\mu$M lasting concentrations of nucleobases to pond environments, which is 1--3 orders of magnitude lower than the experimental limit for reacting these RNA building blocks. Meteoritic delivery of nucleobases offers a higher peak pond concentration of $\sim$10 $\mu$M; however, this source is finite and poses a problem for the chemical evolution of RNA. Organic hazes on the other hand are a constant feeding source to ponds, and this constant replenishment of nucleobase pond concentrations provides the potential for RNA to continuously form and evolve.

Turning off the seepage sink in our models, akin to clogging up rock pores at the base of the pond with amphiphiles or mineral gels \citep{Damer_Deamer2019,Deamer2017}, or biomolecule adsorption onto mineral surfaces \citep{Hazen_Sverjensky2010}, brings both maximum uracil and thymine concentrations up to the $\sim$100 $\mu$M experimental concentration limit after $\sim$8,000 years. The other nucleobases are restricted from reaching 100 $\mu$M in this scenario due to their comparatively high hydrolysis rates; however, in this scenario guanine, adenine and xanthine increase in maximum concentrations to $\sim$5--20 $\mu$M. Turning off seepage does not greatly affect concentrations of hypoxanthine or cytosine. Hydrolysis of organic haze particles may produce additional nucleobases and amino acids within the pond, slightly increasing the concentrations from we simulate here \citep{Neish_et_al2010,Poch_et_al2012}. Nevertheless, we suggest that additional mechanisms of concentration within warm little ponds could further boost these nucleobase concentrations, such as interactions with membrane vesicles \citep{Chen_Walde2010,Himbert_et_al2016}. However, more experimentation is necessary to better understand the degree of concentration that these environments can generate.

Although organic hazes may provide the highest, most stable nucleobase pond concentrations for the origin of RNA on early Earth, estimates of the Hadean haze production rate using the proportionality equation derived by \citet{Trainer_et_al2006} leads to a large model uncertainty (spanning four orders of magnitude). To better understand this uncertainty, we take a different approach to estimate the Hadean haze production rate, using our experimental results. Considering that we produced 72 mg of organic haze particles in our low methane (0.5\%) experiment in 6 days, in a plasma volume of roughly 1000 cm$^{3}$ (1 L), our haze production rate per unit volume is approximately 4.4 kg yr$^{-1}$ m$^{-3}$. The energy density of our cold plasma discharge is $\sim$170 W m$^{-2}$ which is about four orders of magnitude higher than the short wave ($<$ 200 nm) UV flux hitting the top of Earth's atmosphere at 3.9 bya, calculated by \citet{2015ApJ...806..137R}. However, the short wave UV flux was at least an order of magnitude higher at 4.4 bya than at 3.9 bya \citep{Ribas_et_al2005}. Given this information, we might expect the haze production rate per unit volume in the atmosphere at 4.4 bya to be roughly three orders of magnitude lower than our experimental value - which would be 4.4 g yr$^{-1}$ m$^{-3}$. This means a 1 km haze-producing layer in the Hadean atmosphere would produce about 2$\times$10$^{18}$ g yr$^{-1}$ of haze particles at 4.4 bya, which is near the maximum value we calculated using the proportionality equation derived by \citet{Trainer_et_al2006} (9$\times$10$^{17}$ g yr$^{-1}$). We note that the uncertainty of this calculation is probably about two orders of magnitude, given A) the rough flux comparison, B) the uncertainty of the volume of the haze-producing atmospheric layer, and C) the assumption that our plasma dicharge experiment produces the same number of haze particles per energy as short-wave UV. Regardless, this does provide some added validity for the upper bounds of our nucleobase pond concentrations.

\section*{Conclusions}

In this paper, we perform atmospheric simulation experiments to produce solid organic haze particles in early Hadean conditions with high and low methane concentrations. To understand the possibility of organic stockpiling, we also heat up a portion of the high methane experiment to 200 $^{\circ}$C for 1 and 7 days to simulate these haze particles sitting on the potentially hot uninhabitable surface. Then, we use GC/MS/MS to analyze the four haze particle samples for building blocks of life - nucleobases, amino acids, and a few other organics. Finally, we explore organic hazes as a prebiotic source of nucleobases to warm little ponds by modeling their concentrations using a sources and sinks pond model.

The main conclusions are as follows:

\begin{itemize}
\item Organic hazes are a viable source of up to 0.2--1.8 $\mu$M concentrations of the five nucleobases in RNA and DNA to warm little ponds.
\item  Removing pond seepage, possibly representing pore clogging with multilamellar amphiphilic matricies or mineral gels can raise the concentrations of uracil and thymine to the 100 $\mu$M range - the lower limit for the experimental production of nucleotides.
\item Stockpiling biomolecules in organic haze particles on the surface of an uninhabitable Hadean Earth is not a viable route to seed future ponds with nucleobases and most amino acids. Heating many of these biomolecules to 200 $^{\circ}$C reduces their concentration or destroys them beyond detection. Some exceptions are isoleucine/norleucine, alanine, and serine.
\item Decreasing atmospheric methane from 5\% to 0.5\% decreases total nucleobase and amino acid abundances in haze particles by a factor of 4.5 and 1.2, respectively. Total haze mass decreases by a factor of seven, which suggests that there is a roughly linear dependence between total biomolecule influx from hazes and atmospheric methane abundance.
\item Our results suggest that the ideal conditions for the origin of life would be when the Hadean atmosphere is rich in methane and producing lots of organic haze, but not so concentrated in \ce{CH4} that the surface becomes uninhabitable. Finding the exact value for this atmospheric methane concentration will require sophisticated climate modeling; however, our results suggest that 0.5\% may be a reasonable value to consider.
\end{itemize}

We now understand that there are multiple plausible sources of the building blocks of life to warm little ponds, and organic hazes are one of the richest. The main challenge moving forward is to A) demonstrate mechanisms of concentration within the pond, beyond evaporation, and B) demonstrate the production of nucleotides and RNA in these conditions. One hypothesis of interest to the authors is that fatty acid membranes provided the microenvironments necessary for the concentration and reaction of these building blocks of life. We intend to explore this hypothesis in future works.





\section*{Acknowledgments}

We thank the two anonymous referees, whose comments improved this work. B.K.D.P. is supported by the NSERC Banting Postdoctoral Fellowship.

\bibliography{Bibliography}

\beginappendixA

\section*{Appendix}

In Table~\ref{RetentionTimes}, we list the gates, retention time ranges, characteristic ions, and collisional energies for each respective daughter ion used for GC/MS/MS detection of nucleobases, amino acids, and other organics. The gates were developed in a previous work \citep{Sebree_et_al2018}, and have been shifted slightly with the swap of a new GC column for this work.

\begin{table*}[ht!]
\centering
\caption{Gate retention times, characteristic ions and collision energies for each respective daughter ion used for GC/MS/MS detection of the biomolecules in this study. Gates were developed in \citet{Sebree_et_al2018} and shifted slightly with the change-out of a new GC column for this work. Not all of these species were detected in the samples for this work. \label{RetentionTimes}}
\begin{tabular}{lclccc}
\\
\multicolumn{1}{c}{Gate} &  
\multicolumn{1}{l}{Retention time range (min)} & 
\multicolumn{1}{l}{Biomolecule} &
\multicolumn{1}{l}{Parent ion (m/z)} & 
\multicolumn{1}{l}{Daughter ions (m/z)} &
\multicolumn{1}{l}{Collision ener. (eV)}
\\[+2mm] \hline \\[-2mm]
1 & 8.15--9.05 & Ethanolamine & 232 & 147,116 & 10,10\\
2 & 9.05--9.3 & Alanine & 158 & 102,73 & 10,10\\
3 & 9.3--9.7 & Glycine & 218 & 189,147 & 10,10\\
4 & 9.7--9.87 & Sarcosine & 260 & 232,147 & 10,10\\
5 & 9.87--10.23 & $\beta$-Aminoisobutryic Acid & 246 & 147,133 & 10,10\\
6 & 10.23--10.38 & $\beta$-Alanine & 260 & 218,147 & 5,20\\
7 & 10.38--10.8 & Urea & 231 & 173,147 & 15,15\\
 & 10.38--10.8 & Valine/Norvaline & 186 & 130,73 & 12,15\\
8 & 10.8--11.18 & Leucine & 200 & 157,73 & 15,20\\
9 & 11.18--11.58 & Isoleucine/Norleucine & 200 & 157,144,73 & 15,10,20\\
10 & 11.58--12.35 & Proline & 184 & 128,73 & 15,25\\
 & 11.58--12.35 & Uracil & 283 & 169,99 & 30,20\\
11 & 12.35--13.25 & Thymine & 297 & 113,73 & 25,25\\
12 & 13.45--13.87 & Cytosine & 282 & 212,98 & 20,25\\
 & 13.45--13.87 & Methionine & 218 & 170,73 & 10,15\\
13 & 13.87--14.18 & Serine & 288 & 100,73 & 20,20\\
14 & 14.18--14.8 & Threonine & 303 & 202,73 & 20,20\\
15 & 14.8--15.35 & Phenylalanine & 336 & 308,73 & 10,20\\
16 & 15.35--15.65 & Aspartic Acid & 418 & 390,73 & 10,20\\
17 & 15.65--15.87 & Hydroxyproline & 314 & 182,73 & 10,20\\
18 & 15.87--16.04 & Cysteine & 304 & 118,73 & 20,20\\
19 & 16.04--16.4 & Hypoxanthine & 307 & 193,73 & 20,25\\
20 & 16.4--16.73 & Glutamine & 432 & 272,147 & 15,15\\
21 & 16.73--17.15 & Adenine & 306 & 192,73 & 10,25\\
22 & 17.15--18.97 & Lysine & 431 & 300,73 & 10,15\\
23 & 19.15--19.95 & Tyrosine & 466 & 438,147 & 10,15\\
 & 19.15--19.95 & Xanthine & 437 & 363,73 & 20,25\\
24 & 19.95--22.15 & Guanine & 436 & 322,73 & 25,30\\
25 & 22.15--24.35 & Tryptophan & 244 & 188,73 & 10,20\\
\\[-2mm] \hline
\end{tabular}
\end{table*}

\begin{figure*}[!hbtp]
\centering
\includegraphics[width=\linewidth]{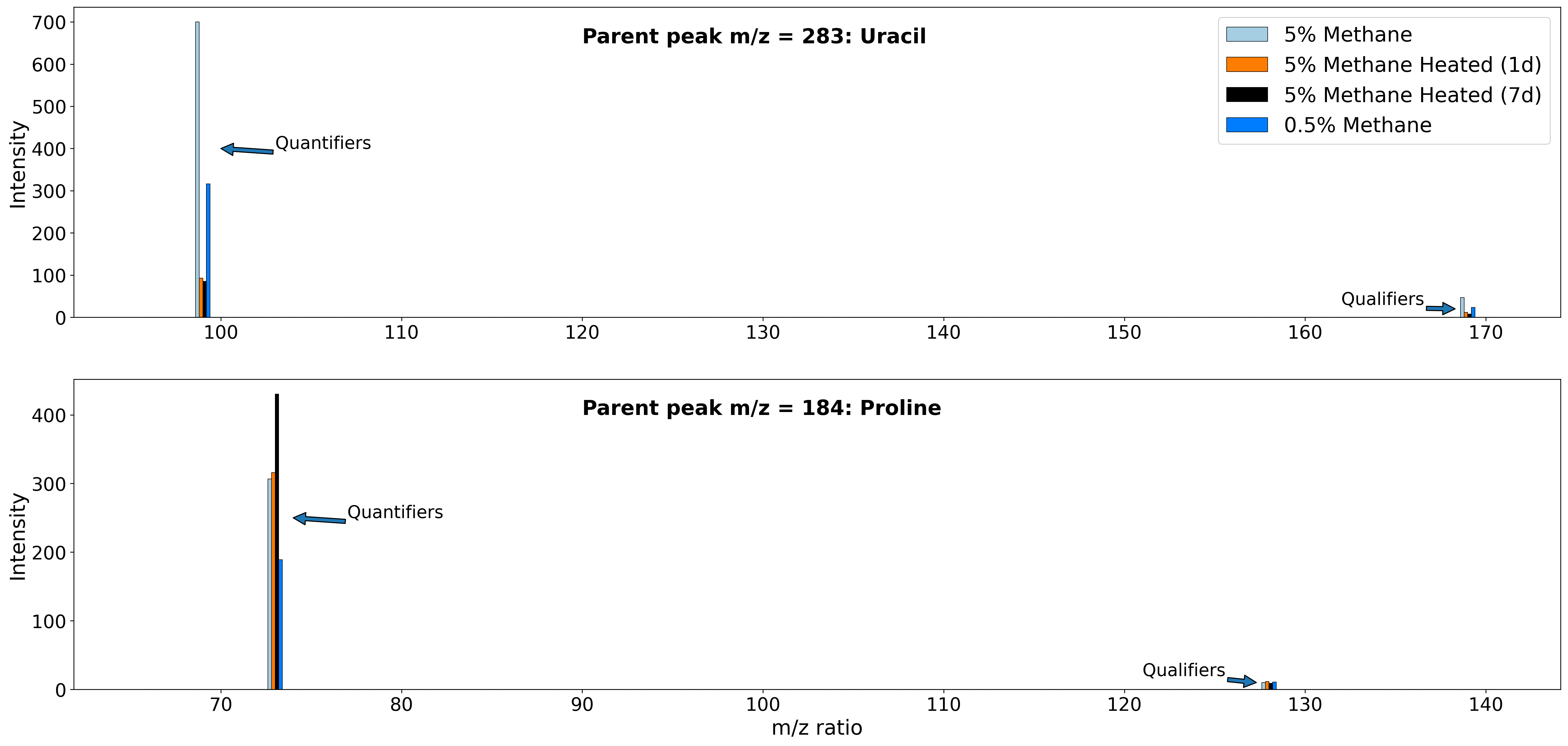}
\caption{GC/MS/MS mass spectra for gate 10 GC peaks integrated over retention time $\sim$11.66-11.77 min, for each of our four haze particle samples. \label{Gate10Peaks}}
\end{figure*}

In Figure~\ref{Gate10Peaks}, we plot the GC/MS/MS mass spectra for the single GC peaks in gate 10 in each of our haze particle samples. Two parent masses are trapped at the first quadrupole for gate 10, m/z = 283 (uracil) and m/z = 184 (proline), and only one daughter qualifier and one daughter quantifier ion are passed along to the third quadrupole for detection. To calculate the relative GC peak percentages, we multiply the GC peak area by the relative quantifier intensities of each species.

\begin{figure*}[!hbtp]
\centering
\includegraphics[width=\linewidth]{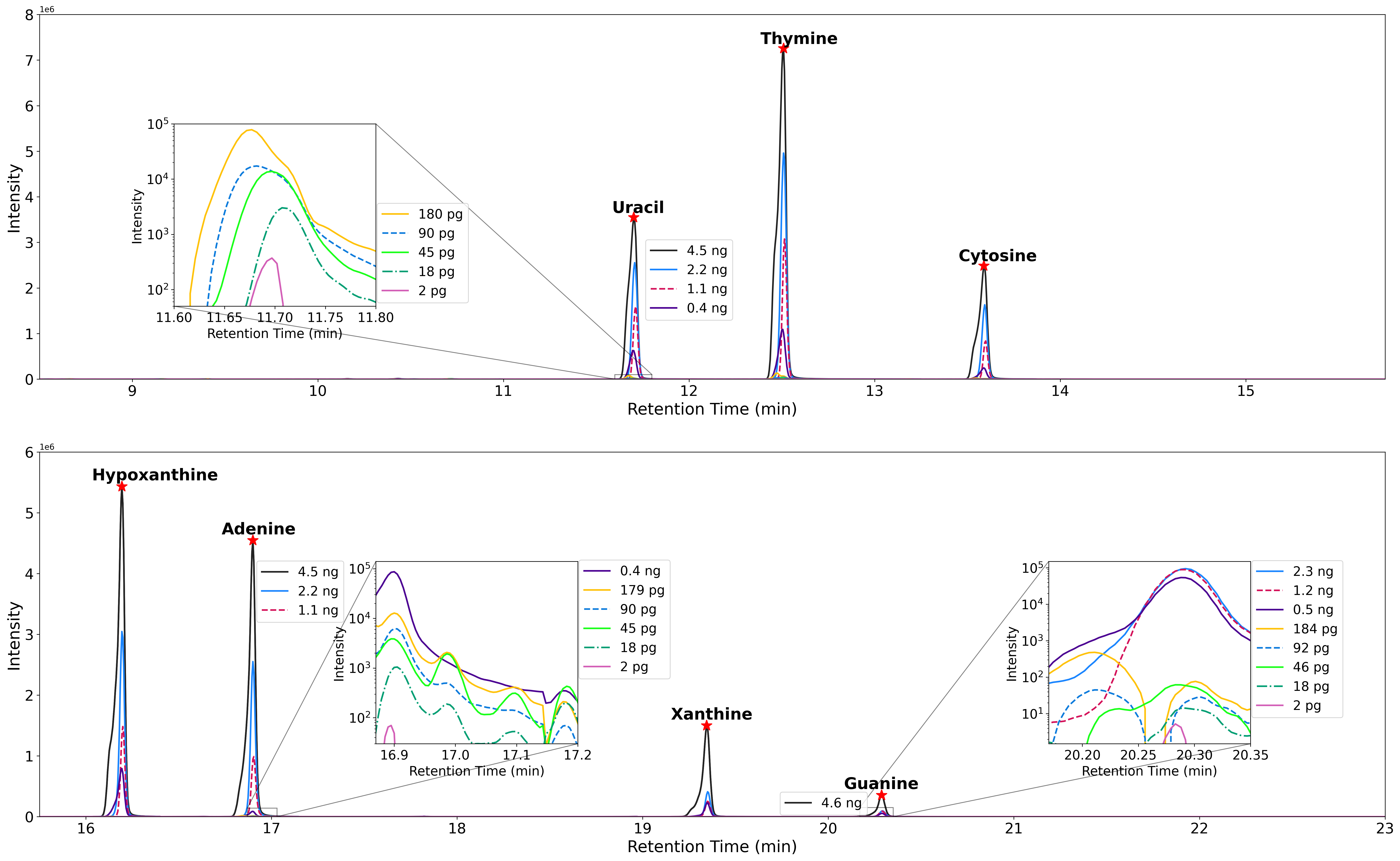}
\caption{GC/MS/MS chromatogram standards for seven nucleobases at nine different sample concentrations/injection volumes (i.e., injection masses). The standards were created in our lab from individual nucleobases purchased from Sigma Aldrich. \label{NucleobaseStandard}}
\end{figure*}

\begin{figure*}[tp]
\centering
\includegraphics[width=\linewidth]{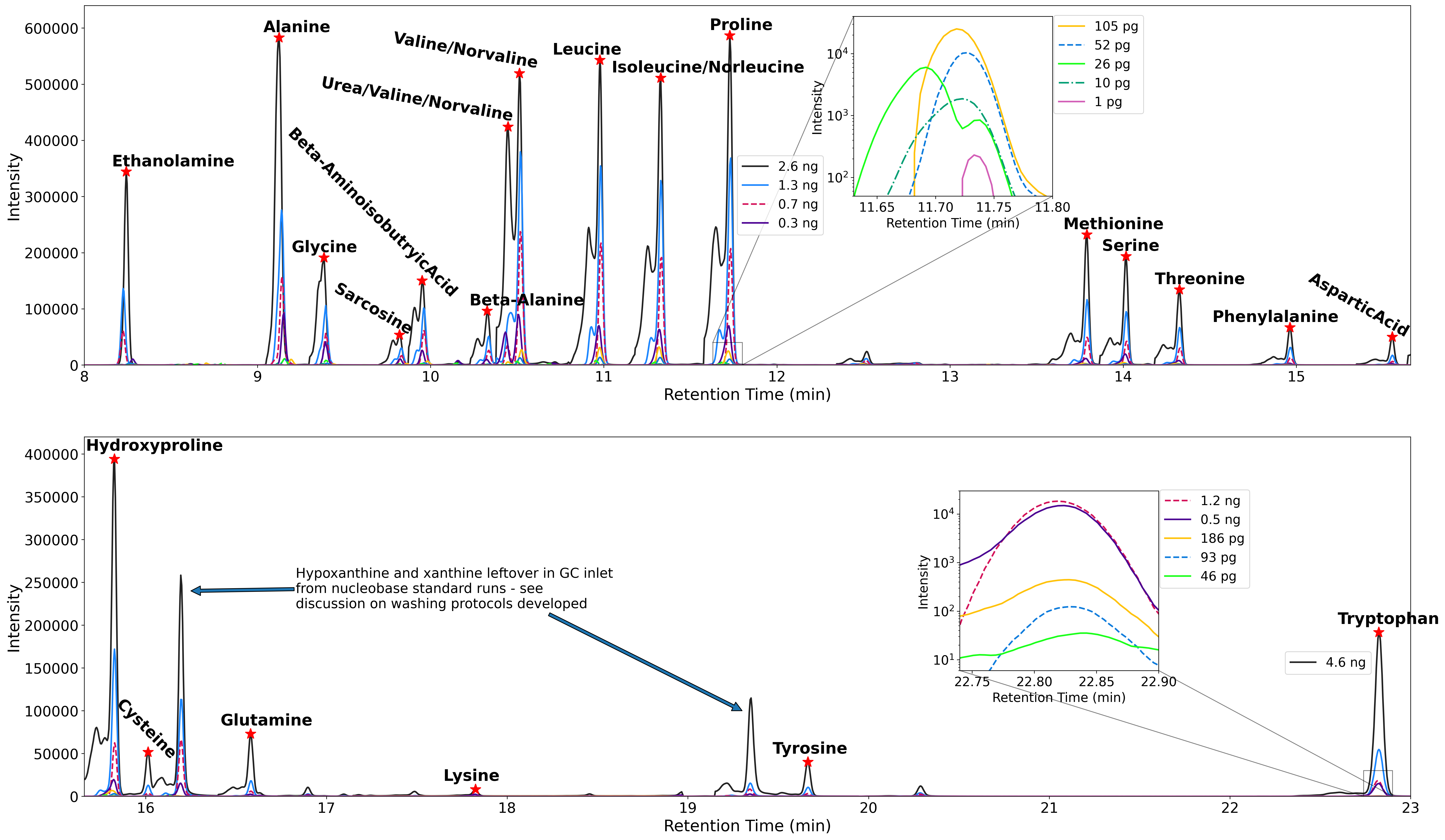}
\caption{GC/MS/MS chromatogram standards for 22 physiological amino acids and other biomolecules at nine different sample concentrations/injection volumes (i.e., injection masses). The complete standard was purchased from Sigma Aldrich. \label{PhysioStandard}}
\end{figure*}

In Figures~\ref{NucleobaseStandard} and \ref{PhysioStandard}, we display the blank subtracted GC standards used for biomolecule detection, and the calculation of calibration curves for biomolecule quantification. We observe that both the guanine and the adenine peaks splits into two peaks at lower concentrations; potentially representing different structural forms. Depending on the sample, one or both peaks are detected in our organic haze particles and the larger peak is used for quantification. The smaller peaks on the left hand side of this highest concentration standard peaks in Figure~\ref{PhysioStandard} are due to peak fronting: an effect that occurs when the sample capacity of the analytical column is exceeded, in our case, due to the injection of too much standard. Only in the case of urea is the highest concentration peak data used in the corresponding calibration curve.

\begin{figure}[tp]
\centering
\includegraphics[width=\linewidth]{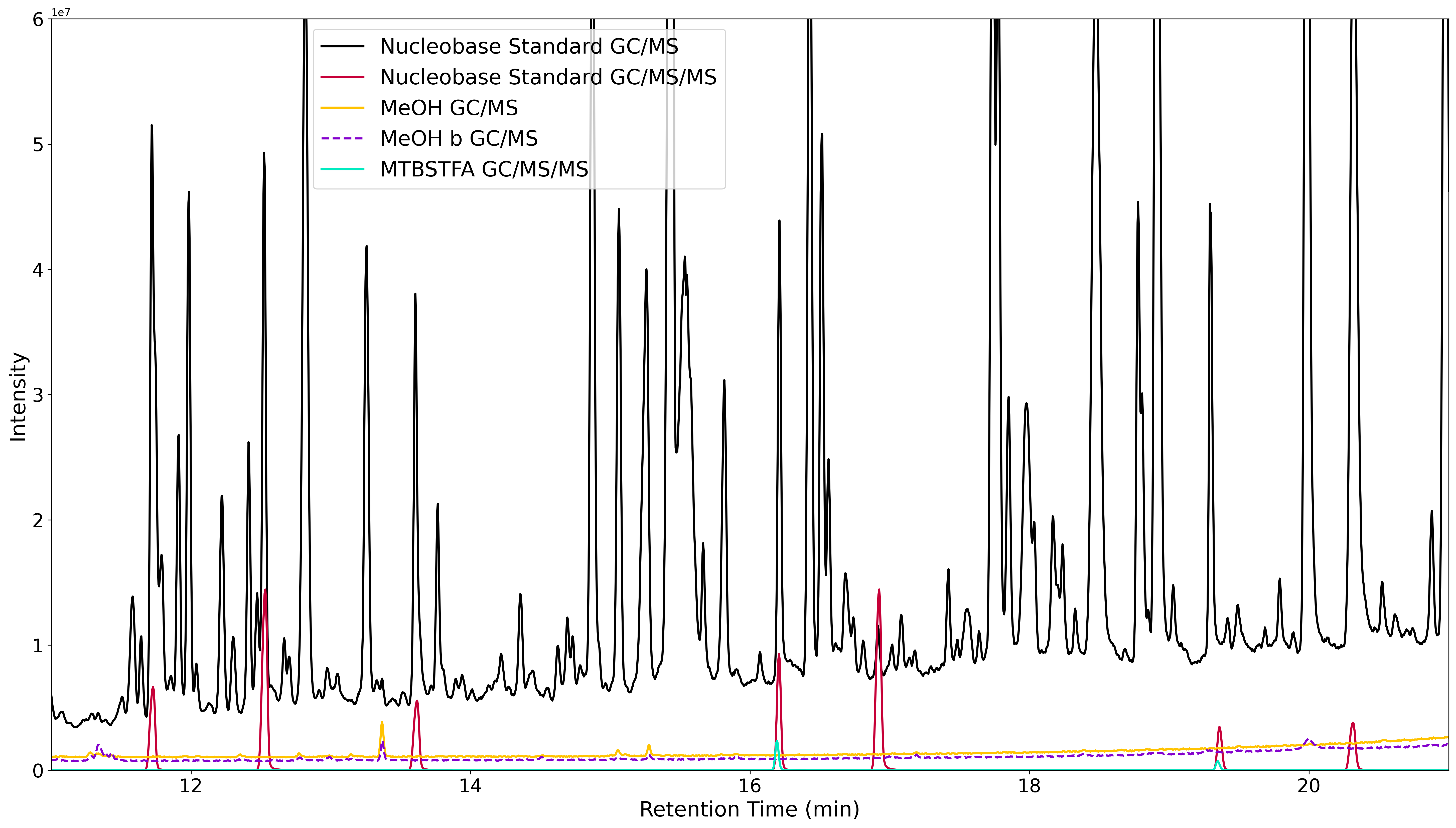}
\caption{GC/MS and GC/MS/MS chromatograms for the nucleobase standard, followed by 2 GC/MS runs of methanol, followed by a GC/MS/MS run of MTBSTFA derivatizer. Nucleobase peaks do not show up in the methanol runs that were run after the nucleobase standard, but do show up when MTBSFTA is run through the inlet, suggesting that underivatized nucleobases stick around in the inlet from the high concentration standard runs. \label{MeOH_spectra}}
\end{figure}

In Figure~\ref{MeOH_spectra}, we display the GC/MS and GC/MS/MS chromatograms for the nucleobase standards that we ran after we swapped out our GC column, inlet and syringe. We followed these runs with two methanol solvent runs in GC/MS mode, and finally, a run with MTBSTFA derivatizer in GC/MS/MS mode. We see no nucleobase peaks in the methanol solvent runs; however, we see decreased nucleobase peaks, particularly hypoxanthine and xanthine, when MTBSFTA is run through the system. This suggests that underivatized nucleobases stick around in the GC inlet from previous high concentration standard runs, and are subsequently washed through the column when derivatizer is run through the inlet.

\begin{figure}[tp]
\centering
\includegraphics[width=0.5\linewidth]{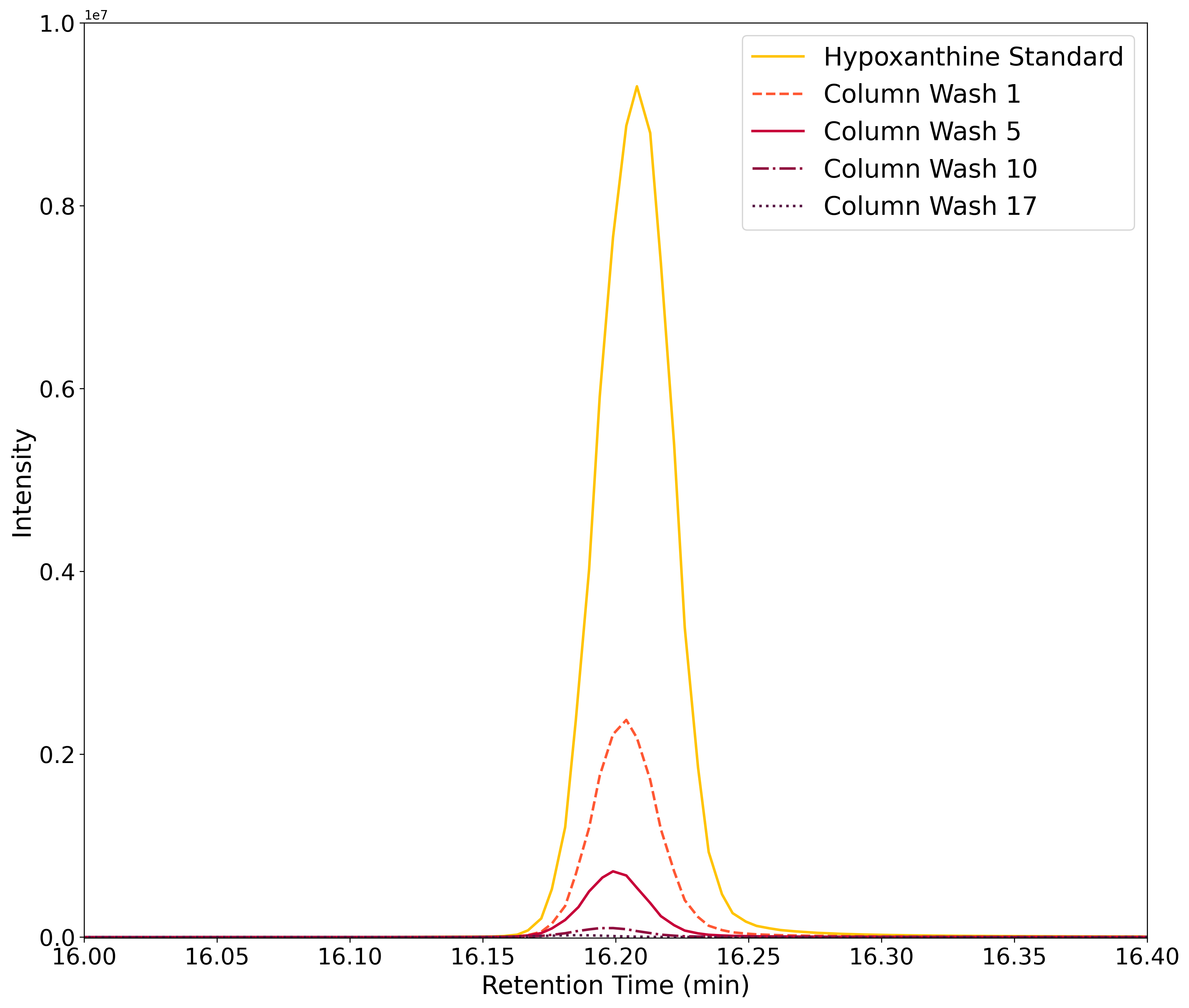}
\caption{GC/MS/MS chromatogram standard for hypoxanthine, followed by 17 GC/MS/MS runs of a procedural blank containing MTBSTFA derivatizer. Only 4 of the wash runs are displayed for clarity. \label{Hypoxanthine_Wash}}
\end{figure}

In Figure~\ref{Hypoxanthine_Wash}, we display the effectiveness of the washing procedure that we developed after discovering that some underivatized biomolecules from our high-concentration standards remained in the GC inlet after each run. Here we see the hypoxanthine peak decrease sequentially after each wash with MTBSTFA derivatizer.

\begin{figure*}[tp]
\centering
\includegraphics[width=\linewidth]{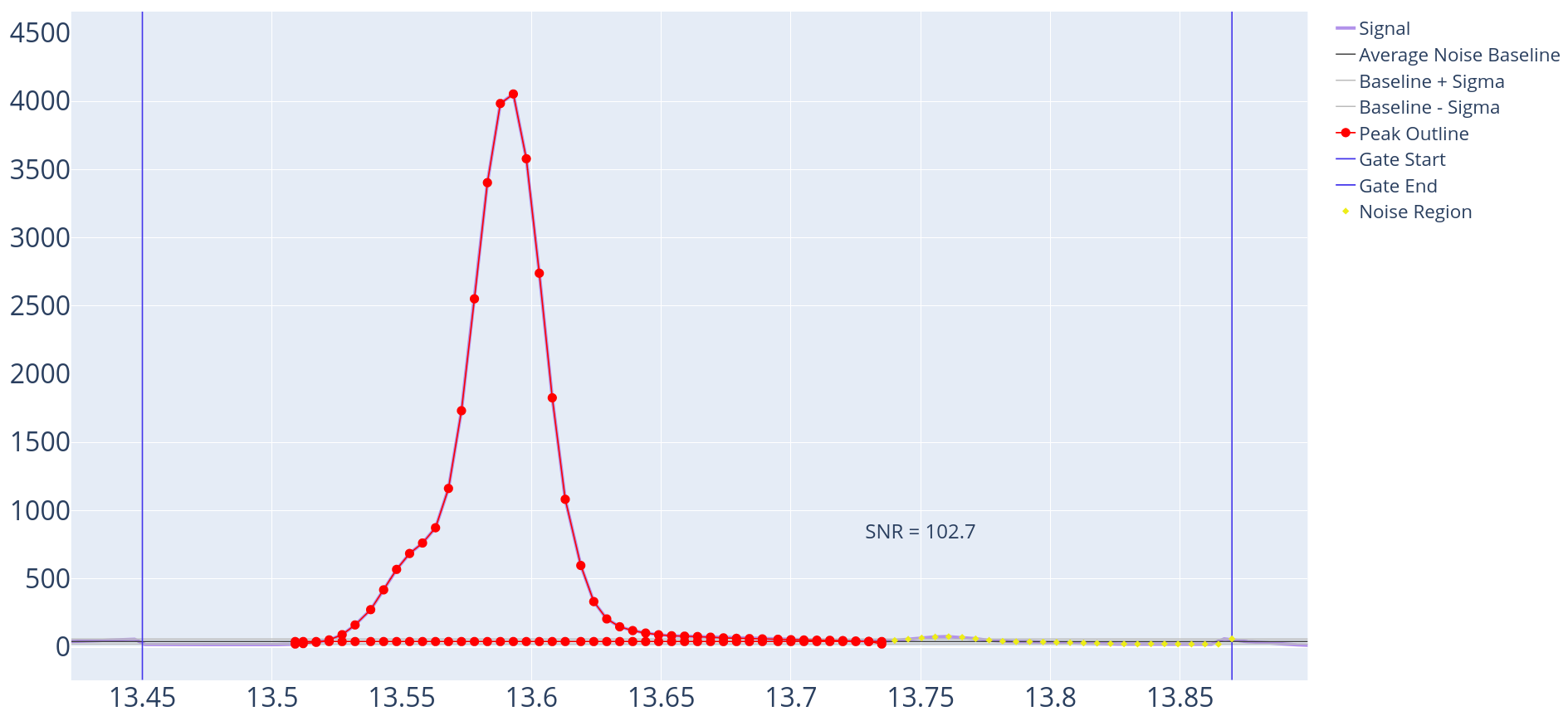}
\caption{A screenshot of the interactive Python code that we developed to find peaks, calculate peak areas, and estimate peak area uncertainty. Noise regions within each gate are manually selected and used to calculate SNR. This peak represents the cytosine detection in our high methane experimental haze sample.\label{Peak_Area_Example}}
\end{figure*}

In Figure~\ref{Peak_Area_Example}, we display a screenshot of the interactive Python code we developed to find retention time peaks within each GC/MS/MS gate, calculate their areas, and estimate the area uncertainties. The noise region within each gate must be selected manually, and is used to calculate the SNR. Peak area uncertainties are calculated by adjusting the average noise baseline by plus or minus one standard deviation.

\begin{figure*}[tp]
\centering
\includegraphics[width=\linewidth]{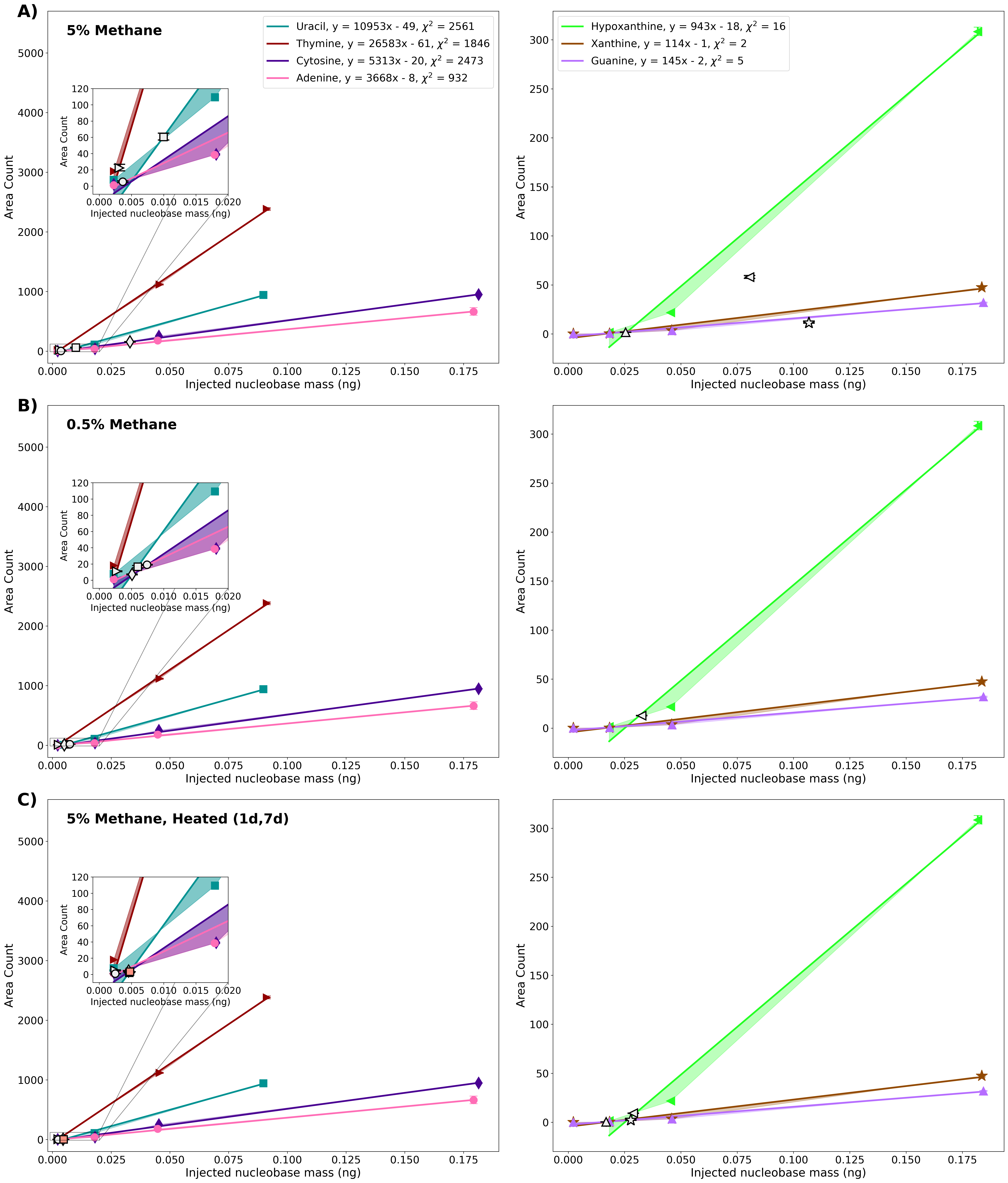}
\caption{Calibration curves calculated for the seven nucleobases. Equations of the lines of best fit and $\chi^2$ values are added to the legend. Uncertainties are the shaded regions calculated as the variance from a model linearly connecting each data point. Peak areas are added for the A) 5\% \ce{CH4} haze particles, 0.5\% \ce{CH4} haze particles, and C) heated 5\% \ce{CH4} haze particles. The uracil data point for the 7 day heated experiment is colored in peach and falls directly on top of the data point for the 1 day heated experiment. \label{Nucleobase_Cal_Curve}}
\end{figure*}

\begin{figure*}[tp]
\centering
\includegraphics[width=\linewidth]{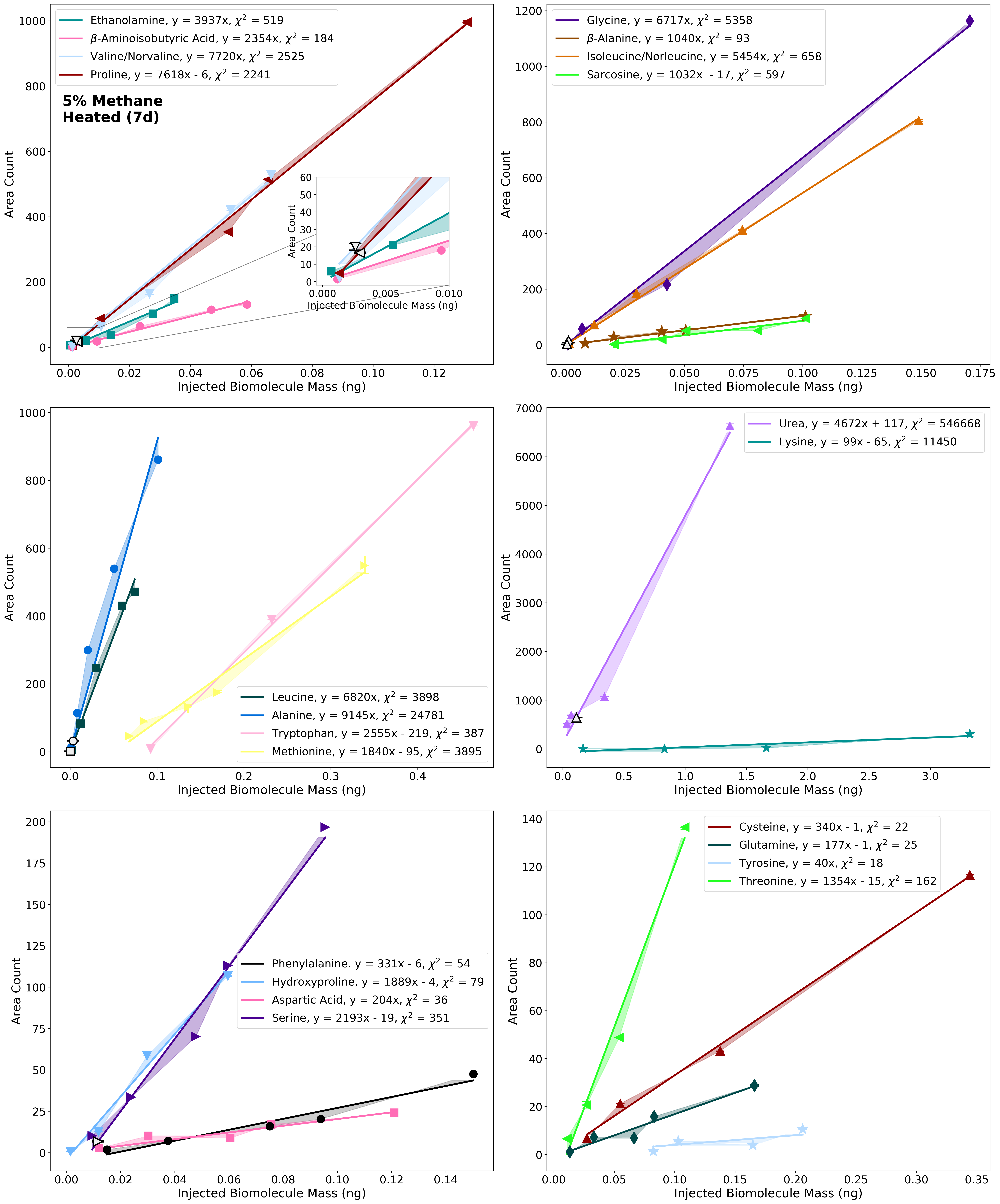}
\caption{Calibration curves calculated for 22 amino acids and other organics. Equations of the lines of best fit and $\chi^2$ values are added to the legend. Uncertainties are the shaded regions calculated as the variance from a model linearly connecting each data point. Peak areas for the 7 day, 200 $^{\circ}$C heated 5\% \ce{CH4} haze particles are added to the plots to demonstrate our quantification calculations. \label{PAmino_Cal_CurveHH7}}
\end{figure*}

In Figures~{\ref{Nucleobase_Cal_Curve}}--{\ref{PAmino_Cal_CurveHH7}}, we display the calibration curves for nucleobases, and amino acids/other biomolecules. To demonstrate quantification, we also insert the peak areas of these biomolecules onto these curves from our high methane experimental haze particles, low methane experimental haze particles, and 1 and 7 day heated high methane experimental haze particles. Uncertainties in the calculated concentrations from our organic haze particle samples take into account both the uncertainty in the GC peak area, as well as the uncertainty in the calibration curve due to am imperfect line of best fit. For example, the uncertainty of valine/norvaline in the 0.5\% organic haze particles is mostly due to the large peak area uncertainty due to the low SNR. On the other hand, the uncertainty of glycine in the 0.5\% organic haze particles is mostly due to the uncertainty in the linear fit at that location on the calibration curve.

\begin{figure*}[tp]
\centering
\includegraphics[width=\linewidth]{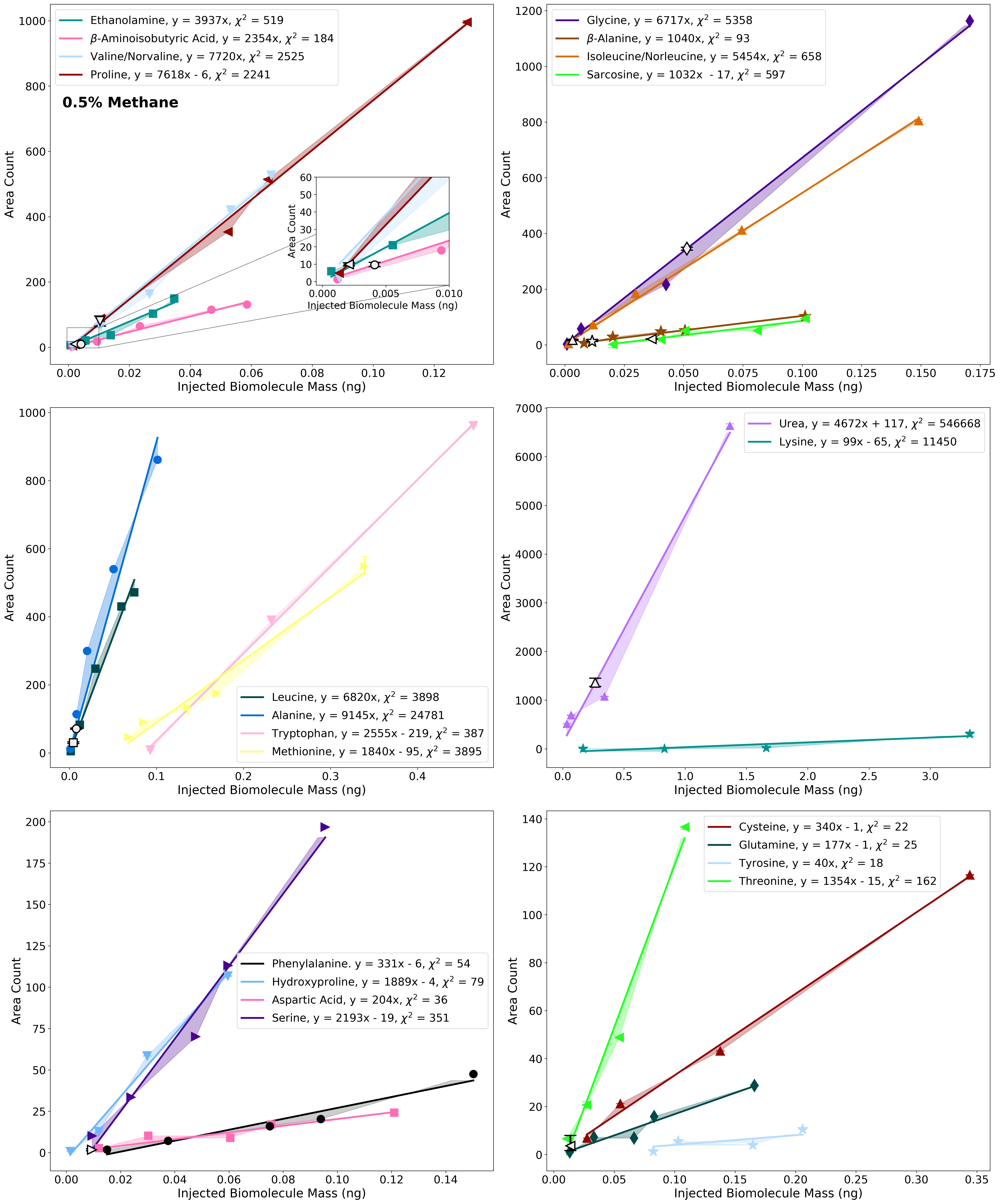}
\caption{Calibration curves calculated for 22 amino acids and other organics. Equations of the lines of best fit and $\chi^2$ values are added to the legend. Uncertainties are the shaded regions calculated as the variance from a model linearly connecting each data point. Peak areas for the 0.5\% \ce{CH4} haze particles are added to the plots to demonstrate our quantification calculations. \label{PAmino_Cal_CurveH05}}
\end{figure*}

\begin{figure*}[tp]
\centering
\includegraphics[width=\linewidth]{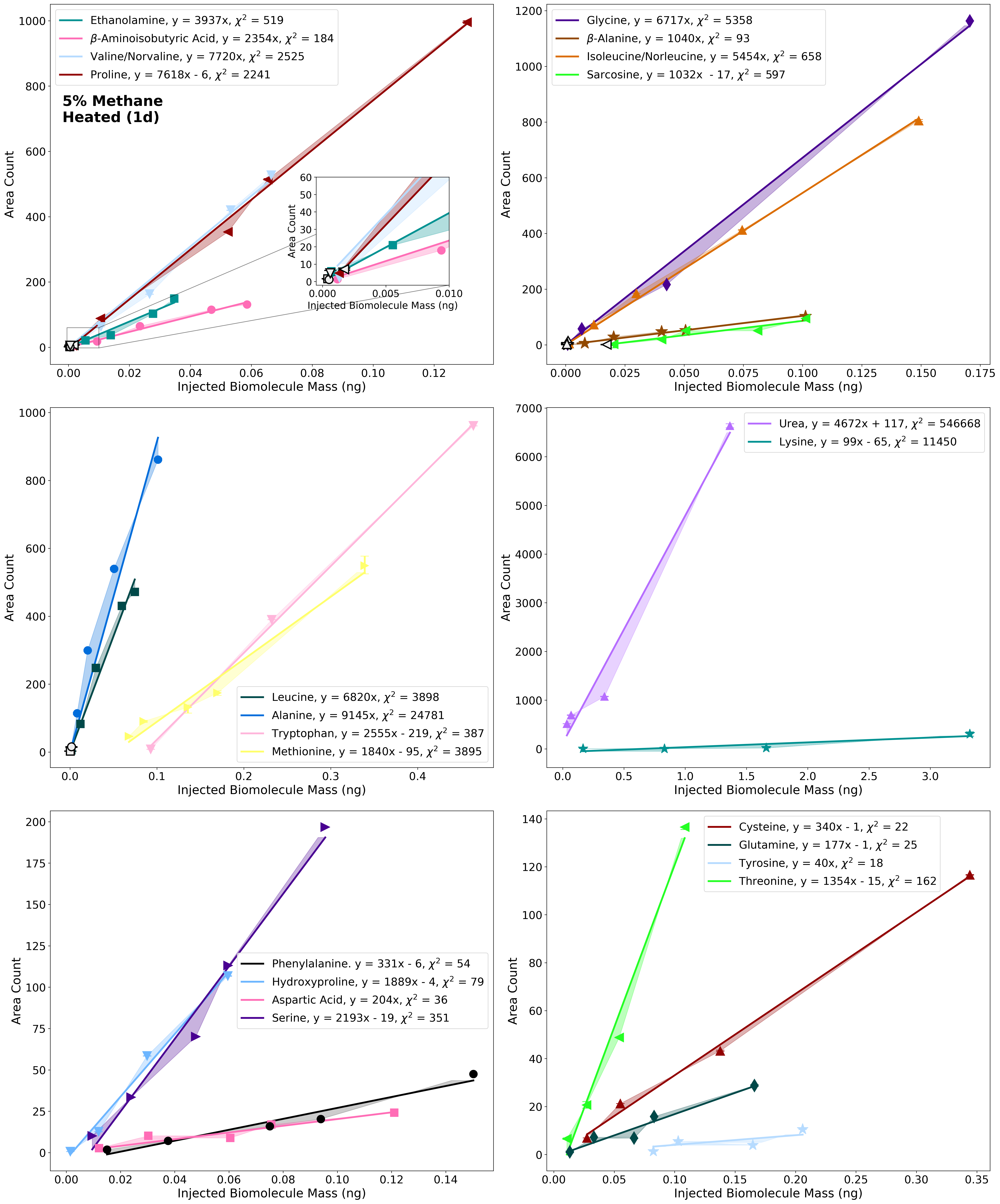}
\caption{Calibration curves calculated for 22 amino acids and other organics. Equations of the lines of best fit and $\chi^2$ values are added to the legend. Uncertainties are the shaded regions calculated as the variance from a model linearly connecting each data point. Peak areas for the 1 day, 200 $^{\circ}$C heated 5\% \ce{CH4} haze particles are added to the plots to demonstrate our quantification calculations. \label{PAmino_Cal_CurveHH1}}
\end{figure*}

In Table~\ref{SNRTable}, we display the lowest detected injection masses and corresponding signal-to-noise ratios (SNR) from the multiple standard runs we performed. SNR are calculated using the peaks from the non-blank-subtracted chromatograms divided by the average noise floor in the corresponding gate. We note that using the blank-subtracted chromatograms leads to erroneous SNR due to negative chromatogram noise intensities from subtraction.

\begin{table}[ht!]
\centering
\caption{Signal-to-noise ratio (SNR) of the lowest detected injection masses for each biomolecule for which we calculated a calibration curve. \label{SNRTable}} 
\begin{tabular}{lcc}
\\
\multicolumn{1}{l}{Biomolecule} &  
\multicolumn{1}{l}{Lowest detected injection masses (pg)} & 
\multicolumn{1}{l}{SNR}  
\\[+2mm] \hline \\[-2mm]
{\bf Nucleobases} & &  \\
Xanthine & 2 & 15 \\
Hypoxanthine & 18 & 9 \\
Cytosine & 2 & 13 \\
Guanine & 2 & 6 \\
Uracil & 2 & 32 \\
Adenine & 2 & 12 \\
Thymine & 2 & 14 \\
{\bf Proteinogenic Amino Acids} & & \\
Tryptophan & 93 & 129 \\
Threonine & 10.8 & 9 \\
Serine & 9.6 & 29\\
Aspartic Acid & 12 & 3.4\\
Alanine & 1 & 30 \\
Proline & 1.3 & 67 \\ 
Glycine & 0.85 & 248 \\
Leucine & 1.5 & 43 \\
Isoleucine/Norleucine & 1.5 & 18 \\
Valine/Norvaline & 1.3 & 7.5\\
Hydroxyproline & 1.5 & 7.3\\
Phenylalanine & 15 & 81 \\
Glutamine & 13 & 100 \\
Methionine & 68 & 707 \\
Cysteine & 27.5 & 211 \\
Tyrosine & 82 & 28\\
Lysine & 166 & 3.6 \\
{\bf Other Biomolecules} & & \\
Urea & 13.6$^*$ & 5\\
$\beta$-Alanine & 8 & 19 \\
Sarcosine & 20 & 44 \\
Ethanolamine & 0.7 & 52 \\
$\beta$-Aminoisobutyric Acid & 1.2 & 7\\
\\[-2mm] \hline
\multicolumn{3}{l}{\footnotesize $^*$ The lowest calibration curve value used is 34 pg.} \\
\end{tabular}
\end{table}

In Table~\ref{LODLOQTable}, we display the injected masses measured in our four haze particle samples, and compare them to the limits of quantification (LOQ) and limits of detection (LOD) roughly determined from our calibration curve data. The LOQ is defined as the injected mass which produces a SNR of $\sim$10, and the LOD is defined as the injected mass which produces a SNR of $\sim$ 3. We can only roughly constrain the LOQ and LOD using the SNRs of the lowest (or two lowest) detected injection masses from our standards (see Table~\ref{SNRTable} for these data).

All of the nucleobase detections in our samples are at or above the upper bound LOQ. This is also the case for most of the proteinogenic amino acids. However, in the case of tryptophan, leucine, isoleucine/norleucine, and valine/norvaline, some of the haze sample measurements are a factor of 1.1--18 below the upper bound LOQ. In the case of these sample data, the detections of leucine, and some detections of isoleucine/norleucine and valine/norvaline have SNR $>$ 10, suggesting these data are above the LOQ. For the detections of tryptophan, isoleucine/norleucine, and valine/norvaline ranging from 3 $<$ SNR $<$ 10, we suggest our measurements as upper limit concentrations in Table~\ref{ConcentrationTable}. Similarly, we suggest upper limit concentrations for the detections of $\beta$-Alanine, sarcosine, ethanolamine, and $\beta$-Aminoisobutyric acid that fall below the upper bound LOQ, and have signals ranging from 3 $<$ SNR $<$ 10.

\begin{table}[ht!]
\centering
\caption{Injection masses measured for each experiment in comparison to limits of detection (LOD), i.e., SNR$\sim$3, and limits of quantification (LOQ), i.e, SNR$\sim$10. The LOD and LOQ are roughly constrained by the lower calibration curve data. Often, only upper bound LOD or LOQ are possible, as the SNR for the lowest standard concentration was often $>$ 3 or $>$ 10, respectively. In the cases where the injection masses for our samples are lower than the upper bound LOD or LOQ, we display the SNR for that measurement. \label{LODLOQTable}} 
\begin{tabular}{lcccccc}
\\
\multicolumn{1}{l}{Biomolecule} &  
\multicolumn{1}{l}{High Methane} & 
\multicolumn{1}{l}{Heated (1d)} &
\multicolumn{1}{l}{Heated (7d)} & 
\multicolumn{1}{l}{Low Methane} &  
\multicolumn{1}{l}{LOD} & 
\multicolumn{1}{l}{LOQ}
\\[+2mm] \hline \\[-2mm]
{\bf Nucleobases} & &  \\
Xanthine & 107 & 28 & - & - & $<$2 & $<$2 \\
Hypoxanthine & 81 & 29 & - & 33 & $<$18 & $\sim$18 \\
Cytosine & 33 & 5 & - & 5 & $<$2 & $<$2 \\
Guanine & 25 & 17 & - & - & $<$2 & $\sim$18 \\
Uracil & 10 & 5 & 5 & 6 & $<$2 & $<$2 \\
Adenine & 4 & 2.5 & - & 7 & $<$2 & $<$2 \\
Thymine & 3 & 2.5 & - & 3 & $<$2 & $<$2 \\
{\bf Proteinogenic Amino Acids} & & \\
Tryptophan & 86 (SNR=3.1) & - & - & - & $<$93 & $<$93 \\
Threonine & 11.5 & - & - & 14 & $<$11 & $\sim$11 \\
Serine & 10 & - & 12 & 9.5 & $<$10 & $<$10 \\
Aspartic Acid & - & - & - & - & $\sim$12 & $<$30 \\
Alanine & 5 & 1.7 & 3.5 & 8 & $<$1 & $<$1 \\
Proline & 4 & 1.7 & 3 & 2 & $<$1.3 & $<$1.3 \\
Glycine & 4 & 1 & 1.1 & 51 & $<$0.85 & $<$0.85 \\
Leucine & 3 & 0.4 (SNR=10) & 0.3 (SNR=10) & 4 & $<$1.5 & $<$1.5 \\
Isoleucine/Norleucine & 0.5 (SNR=5) & 0.5 (SNR=7) & 0.4(SNR=11) & 3 & $<$1.5 & $<$1.5 \\
Valine/Norvaline & 4.5 (SNR=5) & 0.6 (SNR=10) & 2.6 (SNR=30) & 6.5 (SNR=5) & $<$1.3 & $<$11\\
Hydroxyproline & - & - & - & - & $<$1.5 & $<$12\\
Phenylalanine & - & - & - & - & $<$15 & $<$15 \\
Glutamine & - & - & - & - & $<$13 & $<$13 \\
Methionine & - & - & - & - & $<$68 & $<$68 \\
Cysteine & - & - & - & - & $<$28 & $<$28 \\
Tyrosine & - & - & - & - & $<$82 & $<$82 \\
Lysine & - & - & - & - & $<$166 & $<$166 \\
{\bf Other Biomolecules} & & \\
Urea & 8943 & - & 112 & 266 & $<$14 & $<$34 \\
$\beta$-Alanine & 21 & 0.6 (SNR=3.1) & - & 12 & $<$8 & $<$8 \\
Sarcosine & 19 (SNR=7) & 17 (SNR=6) & - & 37 & $<$20 & $<$20 \\
Ethanolamine & 4 & 0.4 (SNR=8) & - & - & $<$0.7 & $<$0.7 \\
$\beta$-Aminoisobutyric Acid & 0.6 (SNR=5) & 0.5 (SNR=6) & - & 4 (SNR=5) & $<$1.2 & $<$9 \\
\\[-2mm] \hline
\end{tabular}
\end{table}

\end{document}